%% file: martini.tex
\documentclass[12pt]{emulateapj}
\usepackage{apjfonts}
\usepackage{natbib}

\shorttitle{AGN IN 1<Z<1.5 CLUSTERS} 
\shortauthors{MARTINI ET AL.} 
\slugcomment{ApJ accepted [22 February 2013]}

\newcommand{\ergs}{erg s$^{-1}$}

\newcommand{\chandra}{{\it Chandra}}
\newcommand{\spitzer}{{\it Spitzer}}

\newcommand{\msun}{M$_\odot$}

\newcommand{\MIT}{3}
\newcommand{\Missouri}{4}
\newcommand{\Davis}{5}
\newcommand{\UFlorida}{6}
\newcommand{\Dartmouth}{7}
\newcommand{\JPL}{8}
\newcommand{\audrey}{9}
\newcommand{\NOAO}{10}
\newcommand{\AZ}{11}
\newcommand{\CfA}{12}
\newcommand{\Monash}{13}

\begin{document} 

\title{The Cluster and Field Galaxy AGN Fraction at z = 1 to 1.5:
Evidence for a Reversal of the Local Anticorrelation Between 
Environment and AGN Fraction} 

\author{Paul Martini\altaffilmark{1,2}
 E.~D.~Miller,\altaffilmark{\MIT}
 M.~Brodwin,\altaffilmark{\Missouri}
 S.~A.~Stanford,\altaffilmark{\Davis} 
 Anthony~H.~Gonzalez,\altaffilmark{\UFlorida}
 M.~Bautz,\altaffilmark{\MIT}
 R.~C.~Hickox,\altaffilmark{\Dartmouth} 
 D.~Stern,\altaffilmark{\JPL}
 P.~R.~Eisenhardt,\altaffilmark{\JPL}
 A.~Galametz,\altaffilmark{\audrey}
 D.~Norman,\altaffilmark{\NOAO}
 B.~T.~Jannuzi,\altaffilmark{\AZ} 
 A.~Dey,\altaffilmark{\NOAO}
 S.~Murray,\altaffilmark{\CfA}
 C.~Jones,\altaffilmark{\CfA} 
 \& M.J.I.~Brown\altaffilmark{\Monash}
} 

\altaffiltext{1}{Department of Astronomy and Center for Cosmology and 
Astroparticle Physics, The Ohio State University, 140 West 18th Avenue, 
Columbus, OH 43210, martini@astronomy.ohio-state.edu}
\altaffiltext{2}{Visiting Astronomer, North American ALMA Science 
Center and University of Virginia, Charlottesville, VA 22903} 
\altaffiltext{\MIT}{Kavli Institute for Astrophysics and Space Research, Massachusetts Institute of Technology, 77 Massachusetts Ave., Cambridge, MA 02139, USA}
\altaffiltext{\Missouri}{Department of Physics and Astronomy, University of Missouri, 5110 Rockhill Road, Kansas City, MO 64110}
\altaffiltext{\Davis}{Department of Physics, University of California, One Shields Avenue, Davis, CA 95616 and Institute of Geophysics and Planetary Physics, Lawrence Livermore National Laboratory, Livermore, CA 94551}
\altaffiltext{\UFlorida}{Department of Astronomy, University of Florida, Gainesville, FL 32611}
\altaffiltext{\Dartmouth}{Department of Physics and Astronomy, Dartmouth College, 6127 Wilder Laboratory, Hanover, NH 03755} 
\altaffiltext{\JPL}{Jet Propulsion Laboratory, California Institute of Technology, Pasadena, CA 91109}
\altaffiltext{\audrey}{INAF - Osservatorio di Roma, Via Frascati 33, I-00040, Monteporzio, Italy} 
\altaffiltext{\NOAO}{NOAO, 950 North Cherry Avenue, Tucson, AZ 85719}
\altaffiltext{\AZ}{Department of Astronomy and Steward Observatory, University of Arizona, 933 North Cherry Avenue, Tucson, AZ 85721} 
\altaffiltext{\CfA}{Harvard-Smithsonian Center for Astrophysics, 60 Garden Street, Cambridge, MA 02138}
\altaffiltext{\Monash}{School of Physics, Monash University, Clayton, Victoria 3800, Australia} 

\begin{abstract}

The fraction of cluster galaxies that host luminous AGN is an important probe 
of AGN fueling processes, the cold ISM at the centers of galaxies, and how 
tightly black holes and galaxies co-evolve. 
We present a new measurement of the AGN fraction in a sample of 13 
clusters of galaxies ($M \geq 10^{14}$ \msun) 
at $1 < z < 1.5$ selected from the \spitzer/IRAC Shallow 
Cluster Survey, as well as the field fraction in the immediate vicinity of 
these clusters, and combine these data with measurements from the literature to 
quantify the relative evolution of cluster and field AGN from the present to 
$z\sim3$.
We estimate that the cluster AGN fraction at $1 < z < 1.5$ is 
$f_A = 3.0^{+2.4}_{-1.4}$\% for 
AGN with a rest-frame, hard X-ray luminosity greater than $L_{X,H} \geq 
10^{44}$ erg/s. This fraction is measured relative to all cluster galaxies 
more luminous than $M^*_{3.6}(z) + 1$, where $M^*_{3.6}(z)$ is the absolute 
magnitude of the break in the galaxy luminosity function at the cluster 
redshift in the IRAC $3.6\mu$m bandpass. The cluster AGN 
fraction is 30 times greater than the $3\sigma$ upper limit on the 
value for AGN of similar luminosity at $z\sim 0.25$, as well as more than an 
order of magnitude greater than the AGN fraction at $z\sim0.75$. 
AGN with $L_{X,H} \geq 10^{43}$ erg/s exhibit similarly pronounced evolution 
with redshift. In contrast with the local universe, where the luminous AGN 
fraction is higher in the field than in clusters, the X-ray and MIR-selected 
AGN fractions in the field and clusters are consistent at $1 < z < 1.5$. This 
is evidence that the cluster AGN population has evolved more rapidly than 
the field population from $z\sim1.5$ to the present. This 
environment-dependent AGN evolution mimics the more rapid evolution of 
star-forming galaxies in clusters relative to the field. 

\end{abstract}


\section{Introduction} \label{sec:intro} 

Numerous lines of evidence suggest that there is co-evolution, and perhaps 
a physical connection, between the growth of supermassive black holes and 
the formation of stars in galaxies. Perhaps the most striking result is 
the similar rate of evolution of the emissivity from Active Galactic Nuclei
(AGN) and star formation from $z\sim2$ to the present 
\citep[e.g.][]{boyle98,franceschini99,merloni04,silverman08b}. 
At the present day, the correlation between the masses of supermassive black 
holes at the centers of galaxies and the velocity dispersions of their spheroids
also supports co-evolution \citep[e.g.][]{ferrarese00,gebhardt00b,tremaine02} 
and may indicate a causal connection. 
Other evidence for a connection between black holes and galaxy growth 
includes that AGN are much more common in the most 
luminous starburst galaxies \citep[e.g.][]{sanders88,veilleux09} 
and that even low-luminosity AGN are more commonly found in galaxies 
with some young stellar populations compared to otherwise similar inactive 
galaxies \citep[e.g.][]{terlevich90,kauffmann03}. 

These observational correlations have fueled a lot of investigation into 
the processes that drive matter to accrete onto supermassive black 
holes, as well as form new stars. The prevalent, theoretical 
framework is that the most luminous AGN and starbursts are triggered 
by major mergers of gas-rich galaxies \citep[e.g.][]{sanders88,barnes91,
hopkins06a}. Numerous other mechanisms have also been proposed to remove 
angular momentum and fuel star formation and black hole growth at lower 
rates, such as large-scale bars, other weakly nonaxisymmetric variations in 
the gravitational potential, minor mergers, disk instabilities, and turbulence 
in the ISM 
\citep[e.g.][]{simkin80,elmegreen98,genzel08,hopkins11}. The observational 
connection between these mechanisms and lower-luminosity AGN is less 
clear \citep[e.g.][]{fuenteswilliams88,mulchaey97a,martini03b}, most likely 
because there are progressively more ways to fuel progressively smaller 
amounts of star formation and black hole growth 
\citep[see][for a review]{martini04c}. 

The distribution of AGN in clusters of galaxies relative to the field provides 
some valuable, additional observational constraints on fueling processes as a 
function of luminosity or accretion rate, as well as the connection between 
black hole and galaxy growth. This is because additional physical processes 
impact the availability and transport of the cold gas that serves as the 
primary fuel source for the central black hole. These processes include the 
removal of cold gas via ram-pressure stripping 
\citep{gunn72}, evaporation by the hot ISM \citep{cowie77}, tidal effects 
due to the cluster potential \citep{farouki81,merritt83a} and other galaxies 
\citep{richstone76,moore96}, and gas starvation due to the absence of 
new infall of cold gas \citep{larson80}. These physical processes 
have been invoked to explain the relative absence of luminous, star-forming
galaxies in clusters, the scarcity of substantial reservoirs of cold gas, and 
the large fraction of relatively quiescent, early-type galaxies 
\citep{gisler78,dressler80,giovanelli85,dressler99}. Observational studies of 
AGN in local clusters have similarly found that luminous AGN are more rare in 
cluster galaxies compared to field galaxies \citep{kauffmann04,popesso06}, 
although less-luminous AGN appear to be present in comparable numbers 
\citep{martini02,miller03a,best05a,martini06,haggard10}. 

The different, or at least additional, physical processes that influence galaxy 
evolution in clusters make the cluster environment very well suited to study 
the co-evolution of supermassive black holes and galaxies. This is because the 
formation and evolution of galaxies has proceeded at a different rate 
in clusters relative to the field. For example, the stars in cluster galaxies 
appear to have an earlier mean formation epoch than field galaxies of similar 
mass \citep[e.g.][]{vandokkum96,kelson97}. Star formation in cluster galaxies 
is also observed to increase rapidly with redshift
\citep{butcher78,saintonge08,haines09,hilton10,tran10,atlee12}. If the 
evolution 
of AGN in clusters traces the evolution of star-forming galaxies in clusters, 
rather than the evolution of star-forming galaxies and AGN in the field, this 
will be strong evidence that AGN and star formation are physically connected, 
and not just a cosmic coincidence. 

The first evidence that substantial numbers of AGN may be present in 
higher-redshift clusters came from the discovery of three AGN in the $z=0.46$ 
cluster 3C295 by \citet{dressler83}, although subsequent spectroscopic surveys 
of other clusters at similar redshifts did not find AGN in large numbers 
\citep{dressler85}. Later \chandra\ observations of many $z>0.5$ clusters did 
find evidence for AGN through the detection of higher surface densities of 
X-ray 
point sources in the fields of these clusters \citep{cappelluti05,gilmour09}, 
although spectroscopic follow-up observations were only obtained in a few cases 
\citep{johnson03,demarco05}. The first quantitative evidence for a substantial 
increase in the cluster AGN fraction with redshift was presented by 
\citet{eastman07}, who compared the fraction of spectroscopically confirmed AGN 
of similar 
X-ray luminosities in low and high-redshift clusters. A larger 
study by \citet{galametz09} further quantified the increase in the AGN 
fraction based on surface density measurements of X-ray, MIR, and radio AGN. 
\citet{martini09} used a spectroscopically confirmed sample to demonstrate that 
the AGN fraction $f_A$ increases as 
$(1+z)^{5.3}$ for AGN above a hard X-ray luminosity of 
$L_X \geq 10^{43}$ \ergs\ hosted by galaxies more luminous than $M_R^*(z)+1$, 
where $M_R^*(z)$ is the absolute magnitude of the knee of the galaxy 
luminosity function at redshift $z$. 
This study included a total of 32 clusters from the local universe to 
$z\sim1.3$, and included data from many previous cluster studies 
\citep{martini06,eckart06,martini07,sivakoff08}. Several more recent studies 
have also identified AGN in high-redshift clusters and groups 
\citep{rumbaugh12,fassbender12,tanaka12}. The rapid rate of AGN evolution 
is quite similar to the evolution of the fraction of star-forming 
galaxies in clusters of $f_{SF} \propto (1+z)^{5.7}$ reported by 
\citet{haines09}, and 
suggests the AGN and star-forming galaxy populations evolve at similar 
rates in clusters, although both power-law indices are uncertain by 
approximately $\pm 2$. 

The evolution of the AGN fraction in clusters of galaxies quantified by 
\citet{martini09} appears to be substantially greater than the evolution of 
the AGN fraction in the field. Work by \citet{alonsoherrero08} and 
\citet{bundy08} demonstrated that the AGN fraction increases by only about a 
factor of two from $z\sim0.5$ to $z\sim1.2$, which is several times smaller 
than the increase for cluster AGN. This relative evolution appears broadly 
consistent with the behavior of star-forming galaxies over the same 
redshift range. \citet{elbaz07} showed that the fraction of galaxies that are 
star-forming is correlated with local galaxy density at $z \sim 1$, which 
is a reversal of the anticorrelation observed in the local universe. 
Nevertheless, a direct comparison between field and cluster surveys is 
complicated because they often employ different selection 
criteria, such as luminosity in some band, or an estimate of the stellar mass, 
and different AGN luminosity limits, to establish their host galaxy and AGN 
samples. These selection criteria are important because the AGN fraction 
above a given luminosity limit depends on stellar mass 
\citep[e.g.][]{heckman04,sivakoff08,aird12} and the evolution 
of the X-ray luminosity function indicates that more luminous AGN were 
proportionally more abundant at higher redshift \citep{hasinger05,barger05}, 
a phenomenon known as ``AGN downsizing.'' 

In this paper we present a new study of a homogenous sample of clusters 
of galaxies at the crucial redshift range of $1 < z < 1.5$ where earlier 
work implied that the fraction of cluster and field AGN would be 
substantially more similar than they are at the present day. 
These clusters were selected from the \spitzer/IRAC Shallow Cluster 
Survey \citep[ISCS,][]{eisenhardt08}, and have uniform visible, 
near-infrared, 
and \spitzer\ observations, deep \chandra\ observations to identify 
luminous AGN, and substantial photometric and spectroscopic redshift data. 
We describe these datasets further in the 
next section. We use these data to uniformly select AGN with X-ray and MIR 
criteria, as described in \S\ref{sec:agn}, and compare the AGN fraction in the 
clusters with the immediate field environment in \S\ref{sec:rad} to 
demonstrate the similarity of the field and cluster AGN fractions in this 
redshift range. In \S\ref{sec:evol} we calculate the cluster AGN fraction 
for this sample and compare it to lower-redshift clusters, then in 
\S\ref{sec:dis} we discuss the relative evolution of the field and cluster 
AGN fraction from the present day to $z\sim3$. We adopt 
$(\Omega_M, \Omega_\Lambda, h) = (0.3, 0.7, 0.7)$ for the cosmological 
parameters. 

\section{Description of the Data} \label{sec:data} 


\input martini.tab1.tex

Our parent cluster sample was selected from the IRAC Shallow Survey 
described by \citet{eisenhardt04}. This survey covers 8.5 deg$^2$ in the 
NOAO Deep Wide-Field Survey in Bo\"otes \citep[NDWFS,][]{jannuzi99} with at 
least 90s of integration time in each of the four IRAC bands. 
These data were supplemented with additional photometry from 
the \spitzer\ Deep, Wide-Field Survey \citep{ashby09}, which added nine 
more 30s exposures across the entire area with IRAC. Deep $B_W$, $R$, and 
$I$ band data from the Mosaic-1 camera on the KPNO 4-m telescope were 
obtained for the NDWFS. Near-infrared images from the FLAMINGOS 
Extragalactic Survey are also available for half of the field 
\citep{elston06}. Deeper, near-infrared data have more recently been obtained 
for the entire field with NEWFIRM (Gonzalez et al. {in preparation}). 

The photometric data were used by \citet{brodwin06} to calculate photometric 
redshifts and redshift probability distribution functions $P(z)$ with an 
empirical template-fitting algorithm. 
The large AGN and Galaxy Evolution Survey 
\citep[AGES,][]{kochanek12} of the Bo\"otes field, together with other 
spectroscopic redshift surveys in this region, were used to create 
training sets and improve the accuracy of the photometric redshifts. 
Based on over 15,000 galaxies with spectroscopic redshifts, 
these photometric redshifts have an uncertainty of $\sigma_z = 0.06(1+z)$ for 
$0 < z < 1.5$. 
The photometric redshift calculations are described in 
detail in \citet{brodwin06}. 

\citet{eisenhardt08} employed a wavelet analysis technique to identify galaxy 
clusters within the Bo\"otes field for the ISCS. The photometric redshift 
distributions $P(z)$ were used to construct weighted galaxy density maps 
as a function of redshift. These density maps in redshift space were then 
convolved with a wavelet kernel and galaxy cluster candidates were identified 
as peaks in the wavelet-smoothed density maps. The significance level 
for each redshift slice was determined with bootstrap resamples of the 
positions and $P(z)$ distributions for the galaxies. \citet{eisenhardt08} 
identified a total of 106 cluster candidates at $z>1$ with this technique 
and estimated that only $\sim10$\% may be due to chance or projection 
effects. A number of these clusters have been spectroscopically confirmed 
to date \citep[][Brodwin et al., in preparation]{stanford05,brodwin06,elston06,eisenhardt08,brodwin11,zeimann12}, where 
spectroscopic confirmation is defined to mean that at least five galaxies have 
redshifts within $\pm2000(1+z_{spec})$ km s$^{-1}$ of the average spectroscopic 
redshift $z_{spec}$ and lie within $R < 2$ Mpc of the cluster center. 

Our study focuses on thirteen clusters that were among the most 
significant from \citet{eisenhardt08}. 
Eleven of these clusters were targeted as
part of a \chandra\ GO program in 2009 to obtain exposures of 30--40 ks 
\citep[they were also included in the larger XBo\"otes survey;][]{murray05,
kenter05}.  The
remaining two clusters have archival data with sufficient exposures to
include in this study. Two other cluster candidates identified by 
\citet{eisenhardt08} also have similar X-ray data, although are not included 
in this analysis. These two clusters are ISCSJ1427.9+3430, which is likely a 
superposition of $\sim 4$ groups along the line of sight, and ISCSJ1429.2+3425, 
which appears to be a close pair of clusters. We do not include 
ISCSJ1427.9+3430 in our 
analysis because it is unlikely to be a massive cluster. We do not include 
ISCSJ1429.2+3425 because the proximity of the cluster pair would add 
uncertainty to our estimate of foreground and background contamination. 
Table~\ref{tbl:sample} lists the basic properties of the thirteen clusters, 
as well as the datasets that we use for this study. All of the clusters 
were observed with the ACIS-I camera, with the exception of ISCSJ1438.1+3414, 
which was observed with ACIS-S. The field of view of these observations were 
$16.8' \times 16.8'$ (ACIS-I) or $8.4' \times 8.4'$ (ACIS-S). In all 
cases the cluster was approximately centered in the field of view. This 
field of view is adequate to encompass the entire angular extent of these 
clusters out to approximately the $r_{200}$ radius, the radius within which the 
cluster is a factor of 200 overdense relative to the average field value, 
with the exception of some area lost to gaps between the ACIS-I chips. 
These data also include substantial coverage out to larger radii, which 
we use to estimate the field population at the cluster redshift. 

We estimated the approximate masses and sizes of these clusters in several 
ways. First, 
\citet{brodwin07} measured the autocorrelation function of 
ISCS clusters and found $r_0 = 19.14^{+5.65}_{-4.56}$ h$^{-1}$Mpc
at $z=1$, which corresponds to a mean cluster mass of $\sim 10^{14}$ \msun. 
Based on \citet{carlberg97}, we estimate that the typical $r_{200}$ radius 
of these clusters is $\sim 1$ Mpc, which corresponds to $\sim2'$ at the 
typical redshift of these clusters. This mean cluster mass estimate, and 
thus the inferred size, has 
been confirmed with more detailed studies of individual clusters. 
\citet{brodwin11} studied ISCSJ1438.1+3414 ($z=1.414$) and ISCSJ1432.4+3250
($z=1.487$) with deep \chandra\ observations and extensive spectroscopic 
observations. This study showed that 
ISCSJ1438.1+3414 has a velocity dispersion of 
$747^{+247}_{-208}$ km s$^{-1}$ based on 17 members, which corresponds to a 
dynamical mass of log $M_{200,dyn}$ $= 14.5^{+0.3}_{-0.7}$ in solar masses. 
The dynamical mass estimate is in very good agreement with the estimate of 
log $M_{200,X}$ $= 14.35^{+0.11}_{-0.14}$ in solar masses from the cluster's 
X-ray luminosity. 
While there is no dynamical mass estimate for the higher-redshift 
ISCSJ1432.4+3250
cluster, the mass estimated from the X-ray luminosity is similar: 
log $M_{200,X}$ $= 14.4 \pm 0.2$ in solar masses. Many of these clusters 
were also included in the weak-lensing study by \citet{jee11} and these 
mass estimates are all above $10^{14}$ \msun. 
As we do not have individual mass estimates for all thirteen clusters, but do 
know they are similar based on their selection, we adopt a projected radius of 
$2'$ for all the clusters. 

\begin{figure*}
\epsscale{1.25}
\plotone{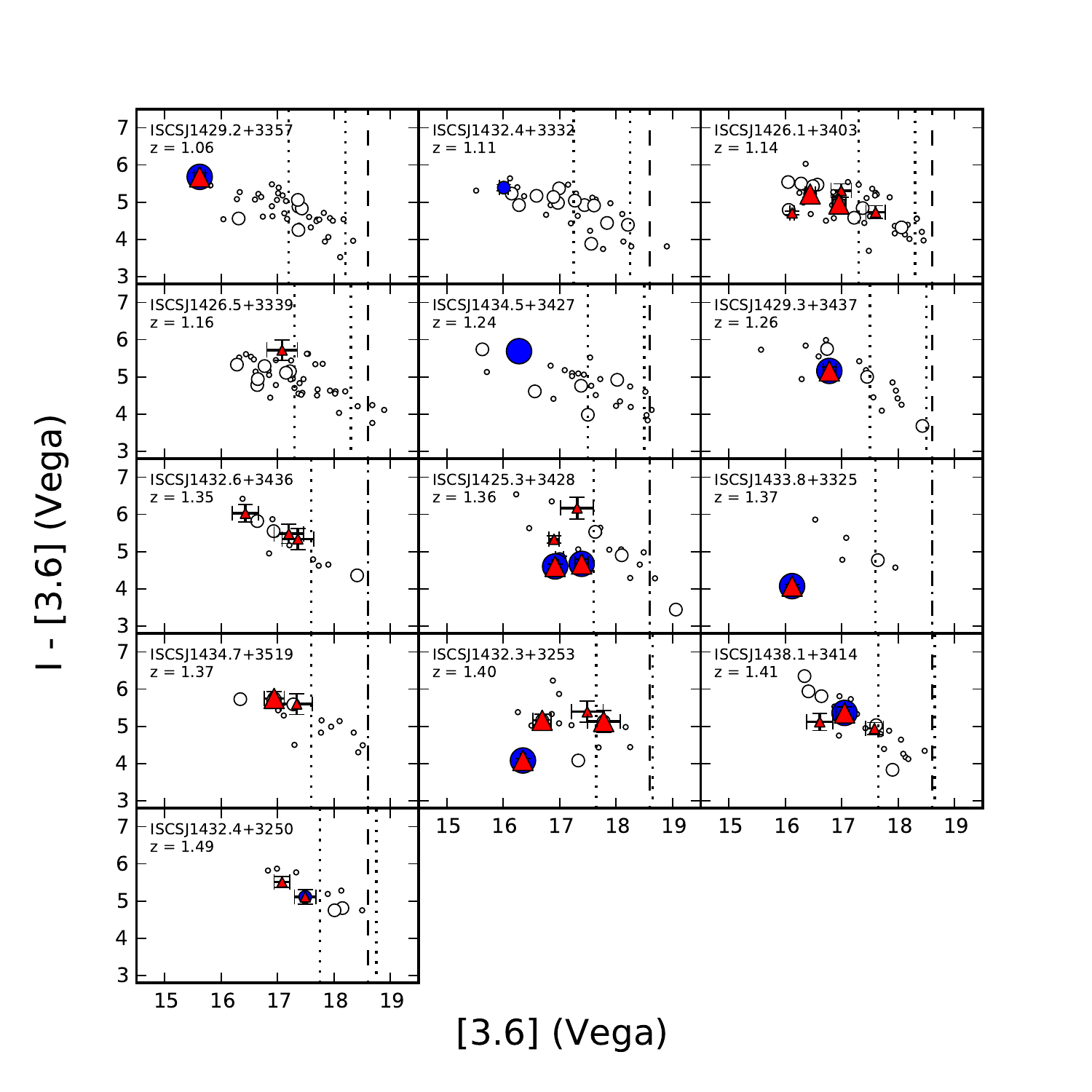} 
\caption{Color-magnitude diagrams for cluster members based on 
spectroscopic ({\it large symbols}) or photometric ({\it small symbols}) 
redshifts. All cluster members are shown that have photometric uncertainties 
less than 0.3 mag in the $I$ and $3.6\mu$m bands and lie within 
$2'$ of the cluster center. Error bars are only shown for X-ray AGN 
({\it blue circles}) and MIR AGN ({\it red triangles}). 
The vertical, dotted lines correspond to the apparent magnitudes of 
$M^*_{3.6}$ and $M^*_{3.6}+1$ at the cluster redshift. Only cluster 
members more luminous than $M^*_{3.6}+1$ are employed in our analysis. 
The vertical, dashed line represents the [3.6] magnitude that corresponds 
to a red galaxy at the [4.5] = 17.8 mag limit employed by \citet{eisenhardt08}. 
\label{fig:cmd}
}
\end{figure*}

There are typically five to fifteen spectroscopically confirmed members in 
each cluster. Because the spectroscopic data are not complete, we supplement 
these data with cluster members based on photometric redshifts for some of our 
analysis. We identify these members based on the integral of $P(z)$ from 
$z-0.06(1+z)$ to $z+0.06(1+z)$, where $z$ is the cluster redshift. Galaxies 
are identified as cluster members if at least $30\%$ of the redshift 
probability distribution is within this range and their position is within a 
projected separation of $2'$ of the cluster center. 
This photometric redshift criterion will include foreground and background 
galaxies. We use a sample of galaxies that satisfy the same photometric 
redshift criterion, but are projected to lie from $2'$ up to $10'$ radius (a 
physical size of $r_{200}$ to $5 r_{200}$), to define a field sample and 
estimate the foreground and background contamination. Many studies have found 
evidence that clusters may impact the surrounding field or `infall' region 
galaxy population at distances up to $5 r_{200}$ 
\citep[e.g.][]{patel09,balogh09}. While we 
estimate the field contamination from this region rather than the true field, 
our results below indicate that this distinction is unimportant for our 
analysis. This is likely because galaxies that have already been affected 
by the cluster do not dominate the surface density of galaxies that satisfy 
the photometric redshift criterion. We also only include galaxies that are no 
fainter than one magnitude below the knee of the luminosity function, 
$M^*_{3.6}+1$ at the cluster redshift. The break in the luminosity function for 
clusters over this redshift range was calculated by \citet{mancone10}. 
The apparent magnitude that corresponds to the break in the luminosity 
function at the redshift of each cluster is listed in Table~\ref{tbl:sample}. 
Figure~\ref{fig:cmd} shows a color-magnitude diagram for each cluster, where 
larger symbols refer to membership based on spectroscopic redshifts and 
smaller symbols to photometric redshifts. The apparent magnitudes that 
correspond to $M^*_{3.6}$ and $M^*_{3.6}+1$ for each cluster are also 
shown on each panel.

\section{AGN Identification} \label{sec:agn} 


AGN in high-redshift clusters have previously been selected based on radio 
emission \citep{johnson03,galametz09,gralla11} and X-ray emission 
\citep{johnson03,eastman07,galametz09,martini09}. These techniques have 
unambiguously determined that luminous AGN are present in high-redshift 
clusters because only black hole accretion can produce such luminous emission 
at these redshifts. In the first subsection below, we describe how we process 
and then analyze our X-ray observations to identify X-ray AGN in clusters and 
the field and 
characterize their properties. MIR selection based on IRAC color selection 
\citep{lacy04,stern05,assef10,donley12} can also effectively identify AGN at 
these redshifts due to the very different SED shape of AGN relative to normal 
and star forming galaxies. This is described further in the following 
subsection. 

\subsection{X-ray AGN} \label{sec:xagn}


The \chandra\ data were reprocessed with CIAO v4.3 and calibration products
from CALDB v4.4.1.  Point sources were identified from a full-resolution
image in the observed 0.5-7 keV energy range using {\tt wavdetect}, a wavelet
source detection tool available in CIAO.  We searched at wavelet scales
between 1 and 32 pixels (0.492 and 15.7 arcsec) to detect a range of source
sizes and to account for the variable PSF across the ACIS field.  A
{\tt wavdetect} threshold of $10^{-6}$ was chosen, which corresponds to 
the likelihood of incorrectly detecting a source at a given pixel. 
For the ACIS-I
observations, only the four ACIS-I detectors were included in the search,
therefore we expect about four spurious detections in each 2048x2048 pixel
field.  One of the clusters (ISCSJ1438.1+3414) was observed with ACIS-S centered
on the S3 detector; we used the same {\tt wavdetect} parameters for this
observation, only including the single detector.  Detections of greater
than 4 net counts were kept in the intermediate source lists, which resulted
in 70 candidate detections of greater than 2-$\sigma$ confidence in a
typical 35 ksec observation.

One cluster (ISCSJ1426.1+3403) has a substantially shorter on-axis \chandra\ 
exposure (11 ksec).  To validate the {\tt wavdetect} results, which were only
performed for the on-axis observation, we included two additional archival
datasets, OBSID 7945 (40 ksec) and 6995 (10 ksec).  These observations were
taken with the cluster very close to the edge of the field of view, but in
opposite directions North and South.  We used acis\_extract 
\citep{broos10}, which searches for sources in multiple overlapping
observations and accounts for non-uniform exposures and PSF sizes in the
searched region.  This procedure resulted in no additional point source
detections within $4'$ of the cluster center.


\input martini.tab2.tex


\input martini.tab3.tex

We cross correlated our cluster member catalogs and the X-ray point source 
catalogs and identified all cluster members with an X-ray source within 
$2''$ of the IRAC position. Eleven of the cluster galaxies brighter than 
$M^*_{3.6} + 1$ are associated with significant X-ray emission. The 
coordinates, redshifts, and apparent magnitudes of these galaxies are listed 
in Table~\ref{tbl:cagn}. 

We calculated the flux and luminosity of each X-ray source in various 
observed and rest-frame bands with a $\Gamma = 1.7$ power-law model, after 
a correction for Galactic absorption. The counts, significance, 
observed-frame 0.5-8 and 2-8 keV fluxes, and rest-frame 0.5-8 and 
2-10 keV luminosities are provided in Table~\ref{tbl:xagn}. All of these 
X-ray sources have rest-frame, 2-10 keV luminosities greater than 
$L_{X,H} > 10^{43}$ erg/s. Only AGN are known to produce point-source 
emission at these luminosities and we consequently classify all these 
galaxies as X-ray AGN. 
Eight of the 11 X-ray AGN have spectroscopic redshifts that confirm they 
are cluster members, including all four with $L_{X,H} > 10^{44}$ erg/s. 
The presence of X-ray emission was not used as a selection criterion to target 
candidate cluster members for spectroscopy; however, two of the 
spectroscopically confirmed AGN were selected based on their MIR 
colors, as described in the next subsection. 


Our X-ray data do not have uniform depth over the area subtended by each 
cluster and field region. The nonuniformity is due to the gaps between 
the ACIS-I detectors and how well the clusters were centered on the detectors. 
We used exposure maps for each cluster to quantify the fraction of the cluster 
and field area with sufficient depth to identify AGN. We identified a 
sensitivity threshold for each exposure map that corresponds to an AGN with 
a rest-frame 2-10 keV luminosity of $10^{44}$ erg/s at the cluster redshift. 
The fraction of the cluster and field regions above this threshold for each 
cluster is used to quantify the X-ray AGN surface density and X-ray AGN 
fraction in the following sections. 

\subsection{MIR AGN} \label{sec:iragn}

\begin{figure*}
\epsscale{1.25}
\plotone{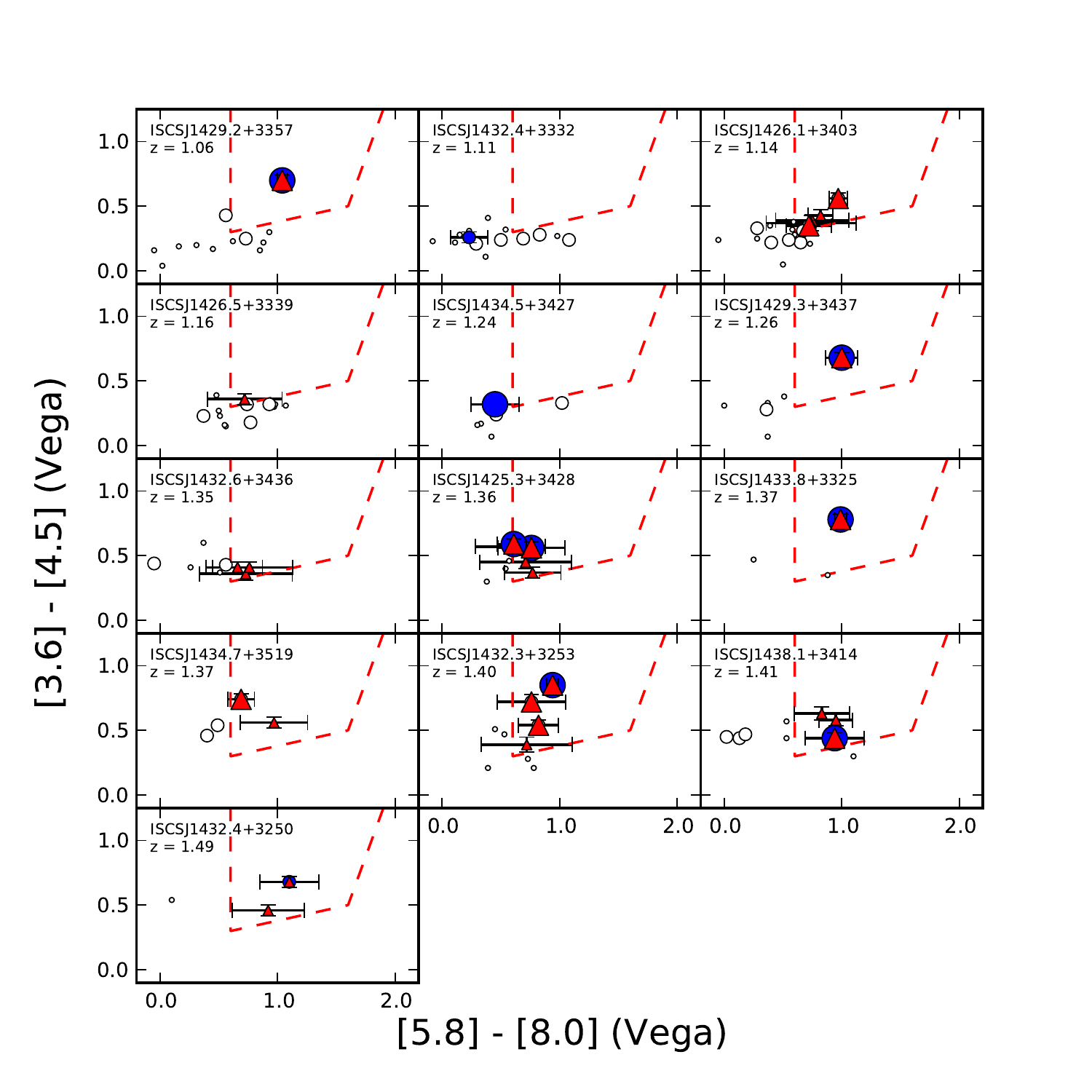} 
\caption{IRAC color-color plots for members of the clusters listed 
in Table~\ref{tbl:sample} ({\it all points}). Error bars are only 
shown for X-ray AGN ({\it blue circles}) and MIR AGN ({\it red triangles}). 
Large symbols correspond to spectroscopically confirmed cluster members. 
Small symbols correspond to photometric redshifts. 
Galaxies are only shown if they are projected to lie within $2'$ of the 
cluster center, have photometric uncertainties less than $0.3$ mag in all four 
IRAC bands, and they are brighter than $M^*_{3.6}(z)+1$, where 
$M^*_{3.6}(z)$ is the [3.6] magnitude of the break in the luminosity 
function at the cluster redshift. 
\label{fig:cc}
}
\end{figure*}

We used our IRAC data to identify candidate cluster AGN with the MIR 
color-selection criteria defined by \citet{stern05} and refer to these 
as MIR AGN. 
Color-color diagrams for the clusters are shown in Figure~\ref{fig:cc}. 
Galaxies were only classified if the photometric uncertainties were less 
than $0.3$ mag in all four IRAC bands. 
There are 27 MIR AGN, of which 12 have spectroscopic redshifts. 
Eight of the 11 X-ray AGN were also identified as MIR AGN, including 
all the X-ray AGN more luminous than $L_{X,H} \geq 10^{44}$ erg/s.  
The incomplete overlap between the MIR AGN and the X-ray AGN is consistent 
with previous studies, which have found similar results for the field 
\citep{hickox09} and in low-redshift clusters \citep{atlee11}. 
Unlike the case for X-ray AGN, J142916.1+335537 in ISCSJ1429.2+3357 and 
J143816.8+341440 in ISCSJ1438.1+3414 were selected for spectroscopic 
observations as cluster MIR AGN candidates. Both of these MIR AGN are also 
X-ray AGN. 

\section{Radial Distributions} \label{sec:rad} 

\begin{figure*}
\plotone{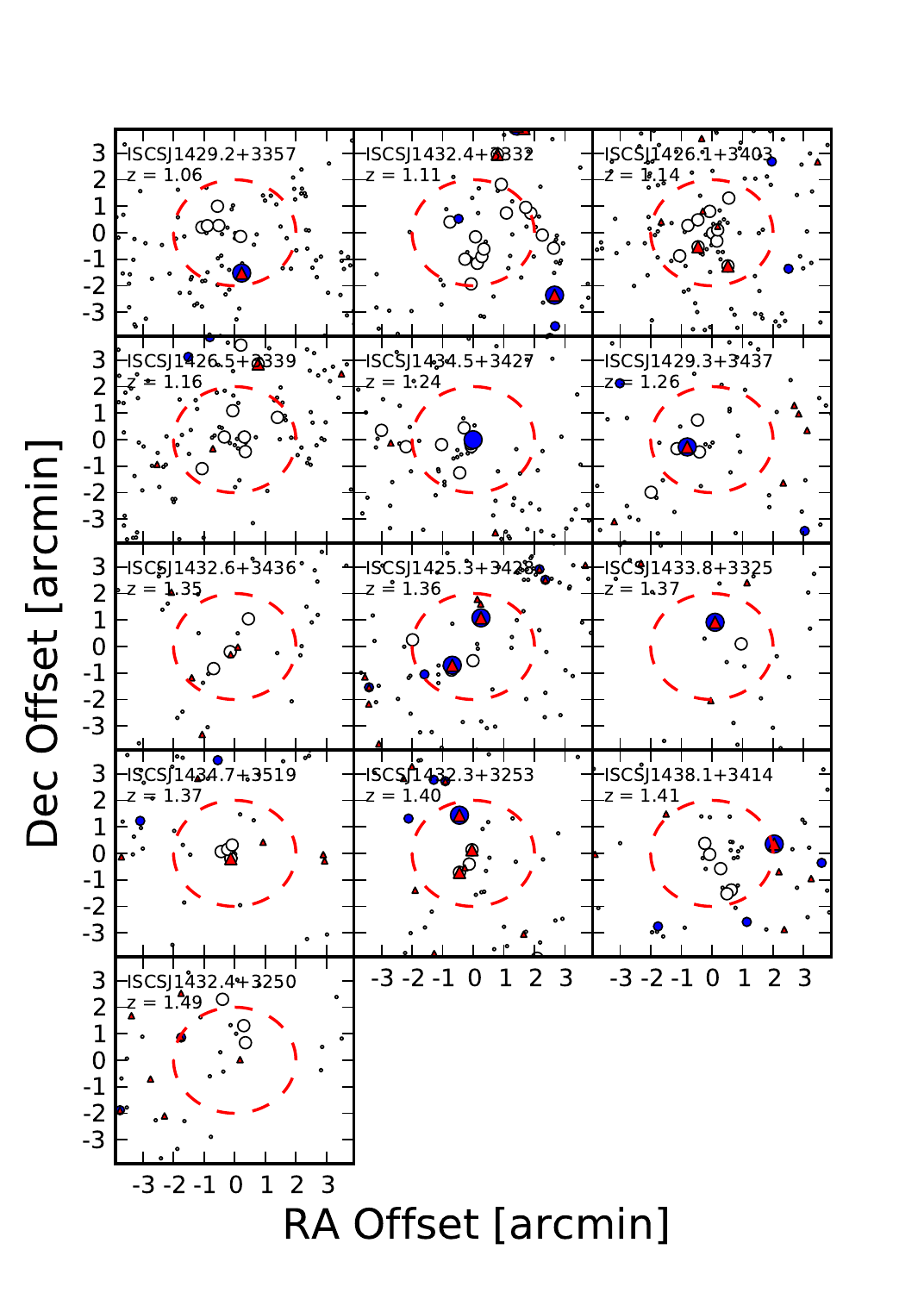} 
\caption{Positions of all cluster galaxies out to a projected distance of 
$2'$ ($\sim r_{200}$, {\it dashed circle}) and field galaxies consistent 
with the cluster redshift within an $8'\times8'$ box centered on the cluster. 
Only galaxies with photometric uncertainties less than 0.3 mag in the 
$I$ and $3.6\mu$m bands are shown. 
Symbols are as in Figures~\ref{fig:cmd} and \ref{fig:cc}. 
\label{fig:pos}
}
\end{figure*}

\begin{figure*}[ht]
\epsscale{1.25}
\plotone{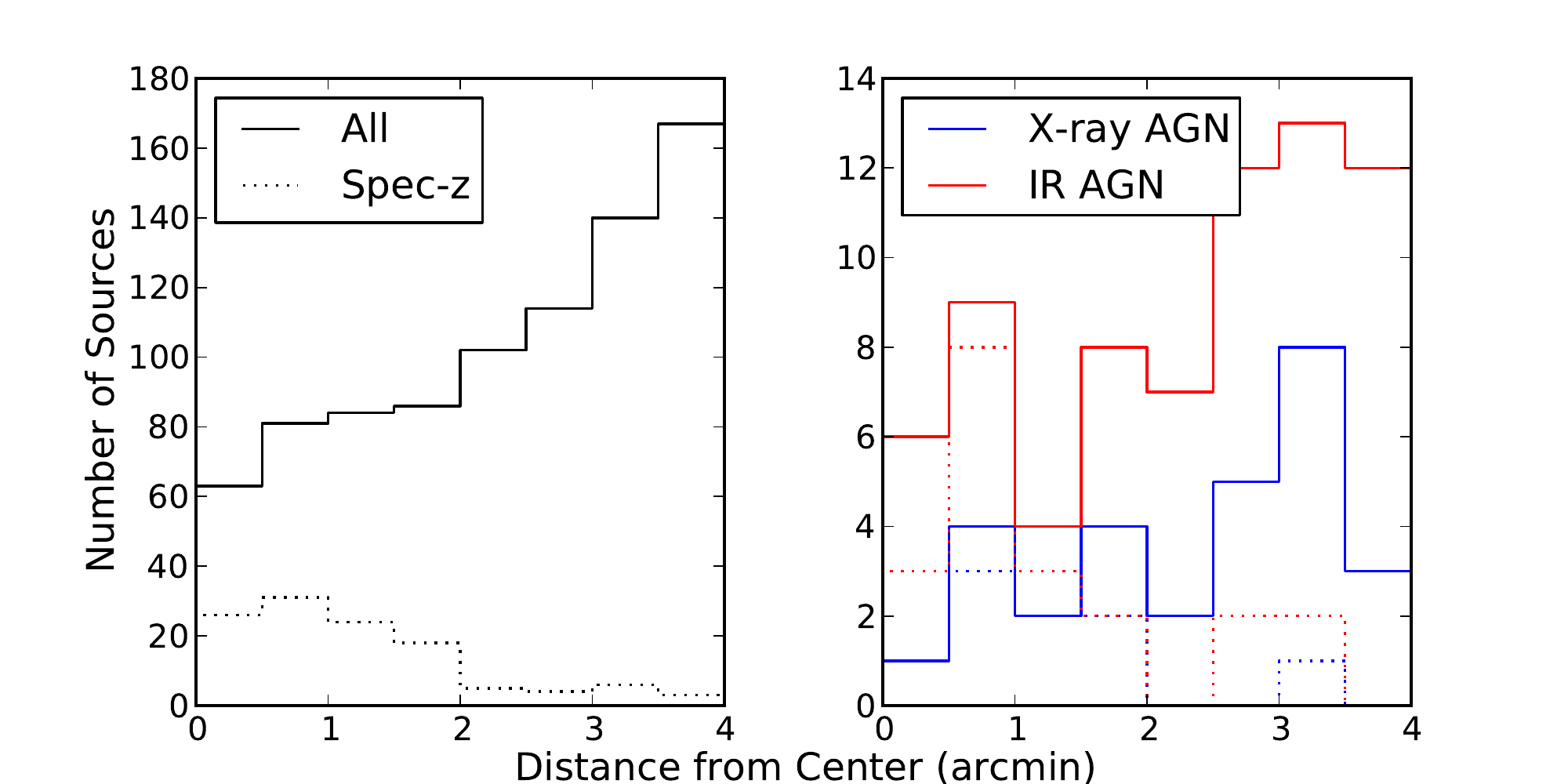}
\caption{Stacked histograms of the galaxies shown in Figure~\ref{fig:pos} as 
a function of clustercentric distance out to $4'$. ({\it Left}) Stack of all 
galaxies with photometric or spectroscopic redshifts ({\it solid line}) and 
only spectroscopic redshifts ({\it dotted line}) consistent with 
cluster membership as a function of projected clustercentric distance. 
({\it Right}) As at left for the X-ray AGN ({\it blue lines}) and MIR AGN 
({\it red lines}) as a function of projected clustercentric distance. 
\label{fig:rad}
}
\end{figure*}

The projected distribution of all galaxies consistent with each cluster's 
redshift, including the X-ray and MIR AGN, is shown in Figure~\ref{fig:pos}. 
The adopted angular size for each of the clusters is $2'$, which is 
approximately the value of $r_{200}$ ($\sim 1$ Mpc) at the redshift of these 
clusters (see \S\ref{sec:data}). Field galaxies consistent with the cluster 
redshift are shown within an $8'\times8'$ box centered on each cluster to 
illustrate the environment in the immediate vicinity of these clusters. 
We use these galaxies, 
including AGN, from $R = 2' \rightarrow 10'$ to measure the surface density of 
field galaxies and quantify the amount of foreground and background
contamination in the cluster sample. A fraction of the cluster and field 
area does not have complete X-ray coverage due to chip gaps and the exact 
placement of the cluster center relative to the center of the ACIS field of 
view. We correct for the non-uniform X-ray coverage as described in 
\S\ref{sec:xagn}. 

As luminous AGN are too rare to study their distribution within individual 
clusters, we stack the cluster catalogs to measure their radial 
distribution. The total number of sources per $0.5'$ bin as a function of 
distance from the cluster center is shown in Figure~\ref{fig:rad}. 
The solid lines correspond to all galaxies that are at the cluster 
redshift. The dotted lines only include galaxies with 
spectroscopic redshifts. The almost complete lack of galaxies with spectroscopy 
outside of $2'$ is because these galaxies were generally not 
spectroscopic targets. Because the spectroscopic coverage is fairly high 
in the clusters, we only use AGN with spectroscopic redshifts to compare to 
AGN in clusters at other redshifts. We use photometric redshifts to estimate 
the total number of cluster galaxies, as well as to estimate the relative AGN 
fractions of X-ray and MIR AGN in clusters and the field. 

\begin{figure*}
\epsscale{1.2}
\plotone{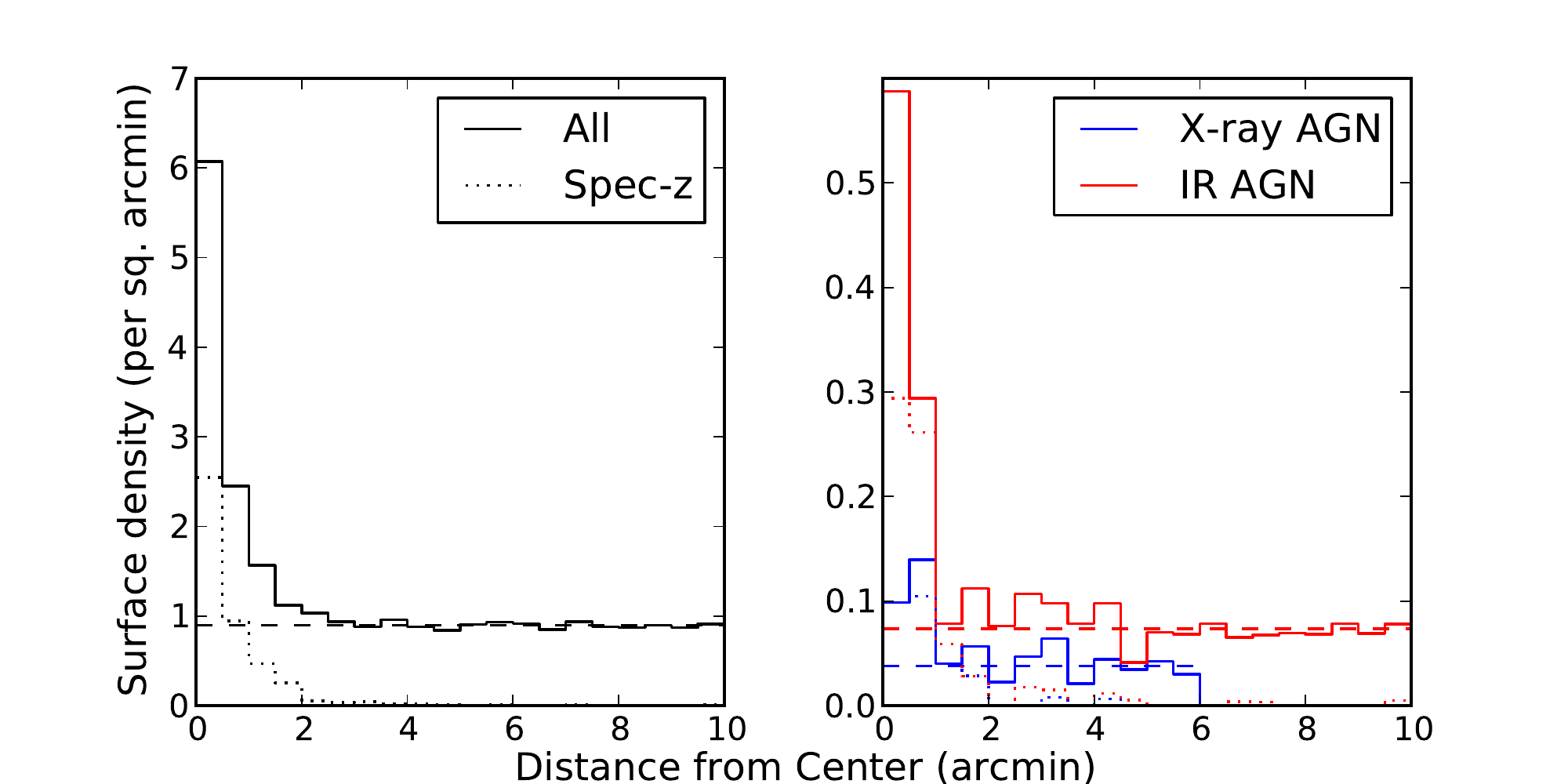}
\caption{
Same as Figure~\ref{fig:rad}, except out to $10'$ from the cluster centers 
and normalized by the area of each annulus to show the density of sources 
per square arcminute. The dashed, horizontal lines correspond to the median 
field surface densities calculated from $2'$ to $10'$ (for all galaxies and 
IR AGN), or from $2'$ to $6'$ (for X-ray AGN) from the cluster center. 
\label{fig:radn}
}
\end{figure*}

Figure~\ref{fig:radn} shows the surface density of all galaxies ({\it left}), 
and X-ray and MIR AGN ({\it right}). The surface densities for all galaxies 
and for MIR AGN are simply the raw counts shown in Figure~\ref{fig:rad} divided 
by the total area of each annulus. The surface density of X-ray AGN is computed 
from the total area in each radial bin that is above the fixed luminosity 
sensitivity threshold described in \S\ref{sec:data}. 
The surface density of galaxies, X-ray AGN, and MIR AGN all asymptote to 
constant values by approximately $2'$ from the cluster center, which is 
consistent with the adopted radius of $2'$ for these clusters. We estimate 
that small uncertainties in our choice of $2'$ for the radius of these 
clusters will not impact our results. 

We use the data from $2' \rightarrow 10'$ to estimate the field surface density 
of all galaxies and MIR AGN, while we use the data from $2' \rightarrow 6'$ 
for all galaxies that have X-ray coverage and X-ray AGN. The radial range 
for the X-ray coverage is smaller because of the size of the X-ray images 
and the deterioration of the point spread function further off axis. 
The surface densities for all galaxies, all galaxies with X-ray coverage, 
X-ray AGN, and MIR AGN are 0.90, 0.99, 0.04, and 0.07 arcmin$^{-2}$, 
respectively. 
We use the field density to calculate the foreground and background 
contamination within the clusters, and then calculate and subtract an estimate 
of the contamination from the total number of galaxies within a projected 
radius of $R = 2'$ to estimate the total number of cluster members, X-ray AGN, 
and MIR AGN in the cluster sample. In the case of X-ray AGN, this includes a 
correction that accounts for the fact that not all the projected cluster and 
field area is above the X-ray luminosity threshold. We estimate
that there are a total 150 galaxies in these thirteen clusters above 
the galaxy luminosity threshold and 136 of these galaxies lie within the area 
above our X-ray luminosity threshold. This surface density extrapolation 
also leads to an expectation of five X-ray AGN and 15 MIR AGN in these 
clusters. These values are consistent with the eight and 12 spectroscopically 
confirmed X-ray and MIR AGN in these clusters (eight AGN are common to both 
samples) listed in Tables~\ref{tbl:cagn} and \ref{tbl:xagn}. 

\begin{figure}[t]
\epsscale{1.25}
\plotone{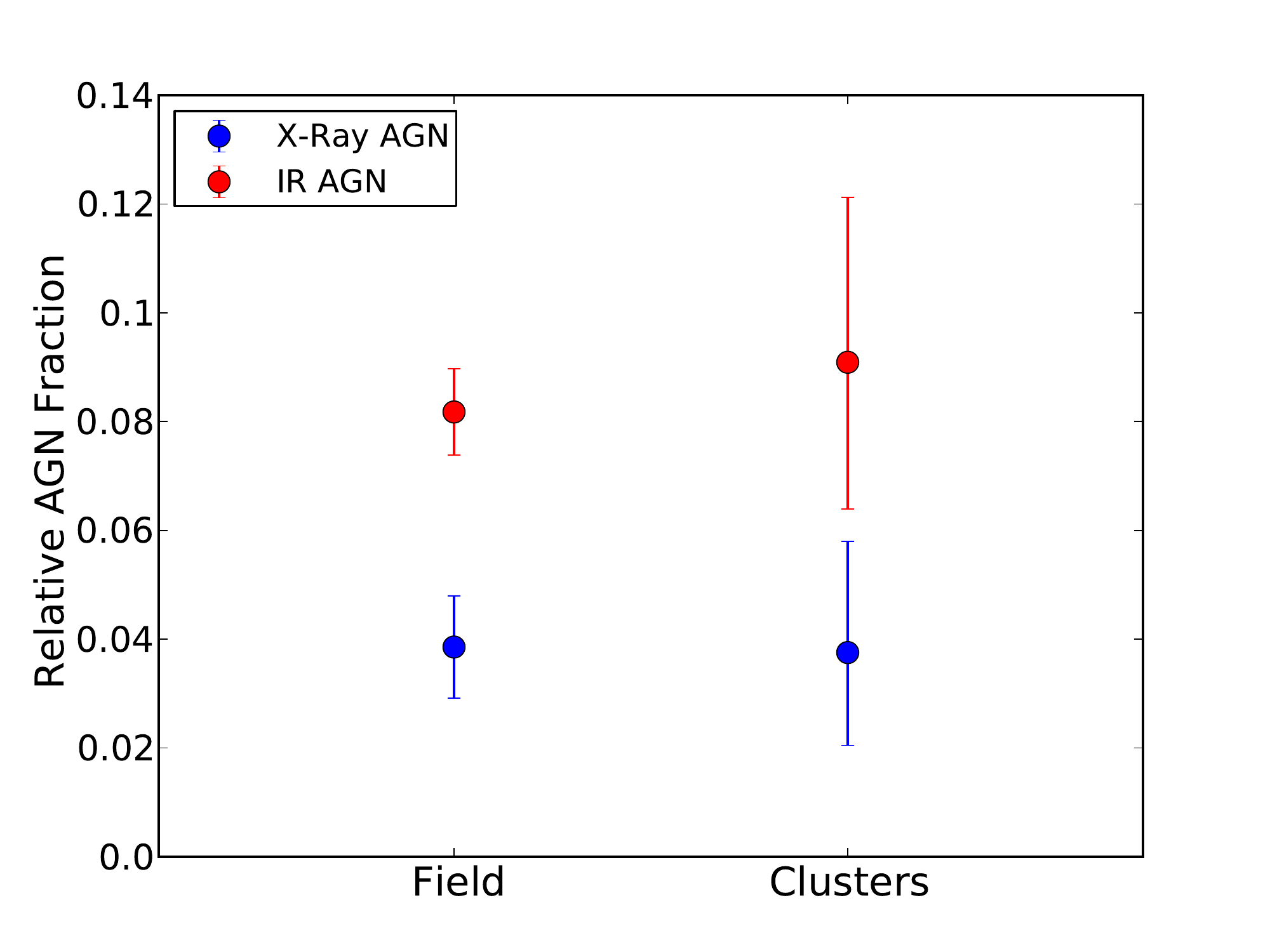}
\caption{Relative fraction of X-ray AGN ({\it blue}) and MIR AGN ({\it red}) 
in the field and cluster samples. The error bars correspond to the 
90\% confidence limits calculated from the relative Poisson uncertainty 
in the number of AGN in each subsample. The field sample is comprised of 
AGN located $2'-10'$ (IR AGN) and $2'-6'$ (X-ray AGN) from each cluster that 
have been selected in an otherwise identical manner to the cluster sample. 
\label{fig:relfrac}
}
\end{figure}

The fractional surface densities of X-ray and MIR AGN in the cluster and 
field samples are consistent with one another. The X-ray and MIR AGN fractions 
in the clusters and the field are shown in Figure~\ref{fig:relfrac}. 
Note that these fractions do not correspond to all AGN above a fixed luminosity 
threshold, but are instead calculated from all AGN above our detection 
threshold in galaxies more luminous than $M^*_{3.6} + 1$. An alternative way to 
express this is to note that the cluster galaxies, cluster galaxies with X-ray 
coverage, X-ray AGN, and MIR AGN correspond to similar overdensities of 
$(\Sigma - \left<\Sigma\right>)/\left<\Sigma\right> = 
1.02, 0.87, 0.82$, and $1.25$, respectively. We discuss the significance of 
the similar cluster and field AGN fractions in \S\ref{sec:dis}. 

There are three biases that influence our estimates. First, the cluster sample 
has been identified via spectroscopic redshifts, or photometric redshifts if 
no spectroscopic data are available, while the field sample has almost 
exclusively been identified via photometric redshifts. As a result of the 
spectroscopic observations, the surface density of foreground and background 
contamination within the clusters is lower than it is outside of them. 
We have compared the number of cluster members within $2'$ based on photometric 
redshifts alone to the number when spectroscopic data are included and find the 
change is $\sim 10$\%, that is we may have slightly underestimated the total 
number of cluster galaxies that form the denominator of the cluster AGN 
fraction (this will not affect the relative AGN fraction in the field and 
clusters). The second bias is that two MIR AGN were selected as targets for 
spectroscopy (see \S\ref{sec:iragn}) and both are also X-ray AGN. These 
constitute only two out of 11 X-ray AGN and two out of 27 MIR AGN, so this 
selection has a minor impact on the relative cluster and field AGN fractions. 
This bias does not impact the numerator of the AGN fractions that are 
derived from spectroscopic redshifts alone. The third bias is that our 
field region only extends to a projected separation of $3 r_{200}$ (X-ray 
AGN) to $5 r_{200}$ (IR AGN). The region out to $5 r_{200}$ is sometimes 
referred to as the ``infall region'' as galaxies at these distances may have 
been ``pre-processed'' by membership in an infalling group or have even already 
passed through the cluster \citep{diaferio01,patel09,balogh09,bahe12} and 
thus may not be representative of the true field population. We expect most 
of the field sample is representative of the true field because the 
photometric redshifts will include many true foreground and background 
galaxies in the field sample; nevertheless, our field estimate may be somewhat 
biased by the infall population. 

\section{Evolution of the AGN Fraction in Clusters of Galaxies} \label{sec:evol}

Eight of the 11 X-ray AGN in these clusters have spectroscopic redshifts that 
confirm their cluster membership. 
All of these X-ray AGN have rest-frame, hard X-ray luminosities 
greater than $L_{X,H} \geq 10^{43}$ erg/s and four (all with spectroscopic 
redshifts) have $L_{X,H} \geq 10^{44}$ erg/s. Our X-ray data are sufficiently 
sensitive that we should have detected all AGN with $L_{X,H} \geq 10^{44}$ 
erg/s at the redshifts of these clusters, with the possible 
exception of AGN in the chip gaps of the ACIS camera. We have examined 
exposure maps of these fields and conclude that chip gaps affect at 
most five of the 350 galaxies that are within the area of these clusters 
and have photometric or spectroscopic redshifts consistent with cluster 
membership. Therefore a negligible fraction of X-ray AGN are missed in the 
chip gaps. 
In \S\ref{sec:rad}, we estimated that there are a total of 136 cluster 
members in galaxies more luminous than $M_{3.6}^* + 1$ with complete 
X-ray coverage (after we correct for foreground and background contamination). 
We therefore estimate that the AGN fraction is 
$f_A(L_{X,H} \geq 10^{44}) = 0.030^{+0.024}_{-0.014}$. 
The uncertainties are Poisson errors that correspond to 
$1\sigma$ Gaussian confidence intervals \citep{gehrels86}. 

\begin{figure}[t]
\epsscale{1.25}
\plotone{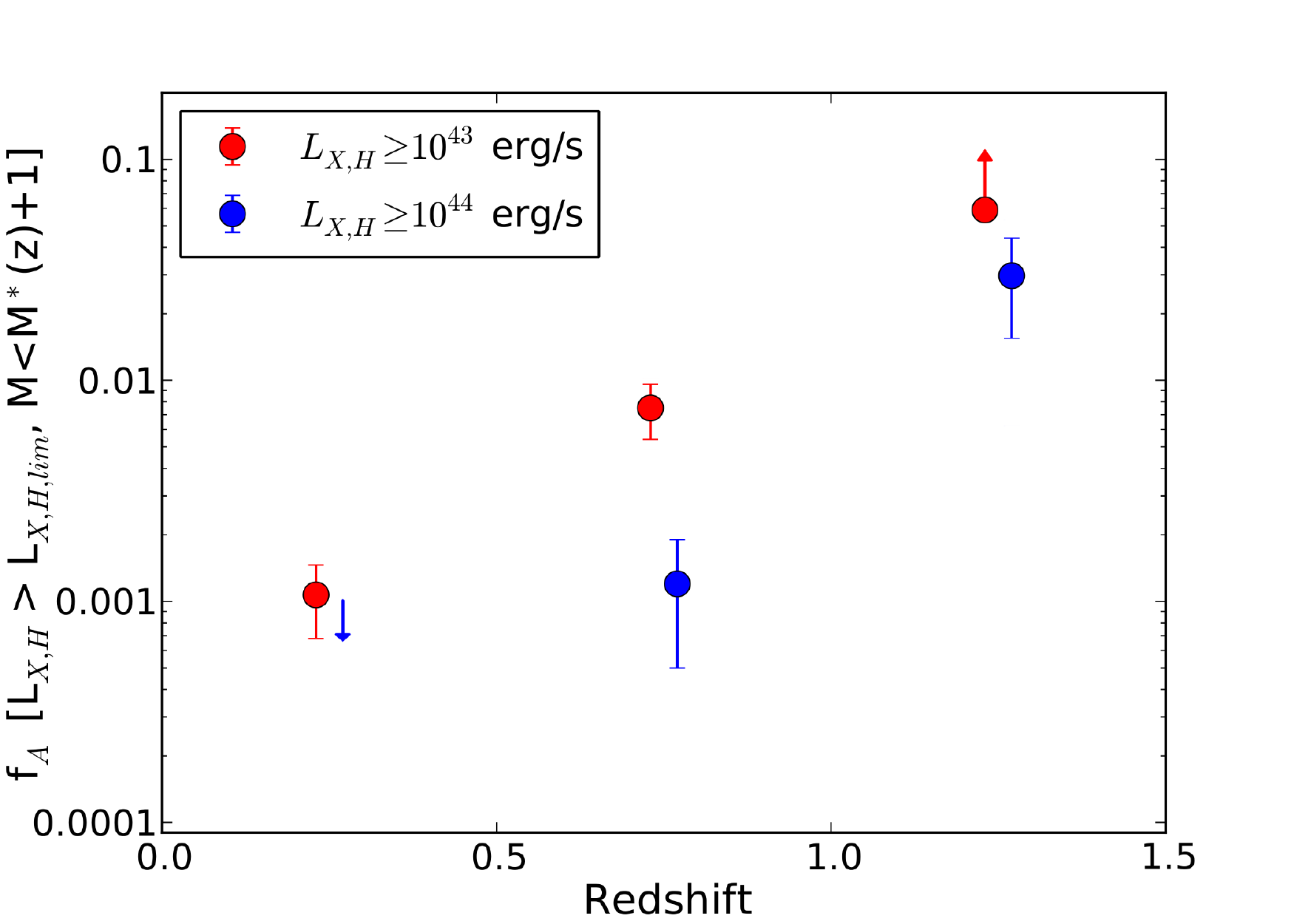}
\caption{Evolution of the X-ray AGN fraction in clusters from $z=0$ to 
$z=1.5$ for hard X-ray luminosity thresholds of $L_{X,H} \geq 10^{43}$ 
erg/s ({\it red}) and $L_{X,H} \geq 10^{44}$ erg/s ({\it blue}). The 
error bars correspond to $1\sigma$ Gaussian errors and the upper 
limit at $z\sim0.25$ corresponds to a $3\sigma$ upper limit. The lower limit 
at $z\sim1.25$ is due to incompleteness. 
\label{fig:agnfracz} 
}
\end{figure}

We are only sensitive to sources as dim as $L_{X,H} \sim 10^{43}$ erg/s in a 
small minority of the clusters. As the luminosity function of cluster AGN is 
not well known at these luminosities, and variations in the spectral energy 
distributions and intrinsic absorption are similarly not well known, we do not 
attempt to model and correct these data to estimate the total number of 
X-ray AGN with $L_{X,H} \geq 10^{43}$. Instead we use our data to establish 
a lower limit on the cluster AGN fraction for sources with 
$L_{X,H} \geq 10^{43}$ erg/s. There are eight spectroscopically confirmed 
AGN in this category (and eleven total), for a lower limit of 
$f_A(L_{X,H} \geq 10^{43}) > 0.059^{+0.029}_{-0.021} (0.082^{+0.032}_{-0.024})$.
The AGN fractions based on the spectroscopically confirmed AGN are shown 
in Figure~\ref{fig:agnfracz} near $z=1.26$, the median redshift of the cluster 
sample (although offset slightly in redshift for clarity). 

Both of these measurements imply that luminous X-ray AGN are more common in 
clusters at $z>1$ compared to lower redshifts, as had been 
indicated by previous work at $z < 1$ \citep{eastman07,galametz09,martini09}. 
\citet{martini09} calculated the AGN fraction and evolution in a similar 
manner to this study for 32 clusters 
at $0.05 < z < 1.27$ (only three at $z>1$) and $L_{X,H} \geq 10^{43}$ erg/s. 
They found that the 
AGN fraction was $f_A(L_{X,H} \geq 10^{43}) = 0.00134$ for $0.05 < z < 0.4$ 
and $0.0100$ for $0.4 < z < 1.27$, where the uncertainty in the measurement
at $z<0.4$ is dominated by the presence of only two AGN in 
the 17 clusters in this redshift range. We have combined the 19 clusters 
at $z<0.5$ presented in \citet{martini09} with the more recent study by 
\citet{haines12} of AGN in clusters at $0.16<z<0.29$ to construct a combined 
sample of 44 clusters and $3869 + 2702 = 6571$ cluster galaxies at 
$z < 0.5$\footnote{Many of the clusters at $0.2 < z < 0.4$ studied by 
\citet{hart09} 
overlap with either the \citet{martini09} or \citet{haines12} sample. Only 
one cluster (A1689) overlaps between \citet{martini09} and \citet{haines12}.}.  
\citet{martini09} found four AGN (two with $0.4 < z < 0.5$) with 
$L_{X,H} \geq 10^{43}$ erg/s (and none with $L_{X,H} \geq 10^{44}$ 
erg/s). \citet{haines12} quote X-ray luminosities in the 0.3-7 keV band, 
rather than 2-10 keV. We used the ratio of the 0.3-7 keV band to the 2-10 keV 
band for a $\Gamma = 1.7$ power law to scale their results to 2-10 keV. This 
yields three AGN in their sample with $L_{X,H} \geq 10^{43}$ erg/s and none 
above $10^{44}$ erg/s.  The combined sample has AGN fractions of 
$f_A(L_{X,H} \geq 10^{43}) = 0.00107^{+0.00057}_{-0.00039}$ and 
$f_A(L_{X,H} \geq 10^{44}) < 0.00101$ (a $3\sigma$ upper limit). This point and 
upper limit are shown in Figure~\ref{fig:agnfracz} near $z=0.25$. 
At intermediate redshifts of $0.5 < z < 1$, \citet{martini09} found 13 AGN with 
$L_{X,H} \geq 10^{43}$ erg/s, including two with $L_{X,H} \geq 10^{44}$ erg/s, 
in ten clusters and an estimated population of 1734 galaxies. The AGN 
fractions are $f_A(L_{X,H} \geq 10^{43}) = 0.0075^{+0.0027}_{-0.0021}$ and 
$f_A(L_{X,H} \geq 10^{44}) = 0.0012^{+0.0015}_{-0.0007}$. 
These points are shown in Figure~\ref{fig:agnfracz} near $z=0.75$. 
Note the luminosity thresholds employed to define the galaxy samples in 
these studies are similar to ours: \citet{martini09} adopted a threshold of 
$M_R^*(z) + 1$ and \citet{haines12} adopted $M_K^* + 1.5$. 

The evolution in the AGN fraction corresponds to a factor of at least 
45 for $L_{X,H} \geq 10^{43}$ erg/s AGN from $z\sim0.25$ to $z\sim1.25$ and 
a factor of at least 30 for $L_{X,H} \geq 10^{44}$ erg/s AGN. In the former 
case, the increase is a lower limit due to incompleteness at $z>1$. 
In the latter case the increase is a lower limit due to the absence of 
any AGN in this luminosity range in the 44 clusters that comprise the 
low-redshift sample. For the two 
luminosity and redshift ranges without limits on the fractions, the evolution 
in the 
AGN fraction is also pronounced. For $L_{X,H} \geq 10^{43}$ erg/s AGN, 
the fraction increases by a factor of seven over the range 
$0.25 < z < 0.75$. For  $L_{X,H} \geq 10^{44}$ erg/s AGN, 
the fraction increases by a factor of 25 over the range 
$0.75 < z < 1.25$. This substantial increase is in good 
agreement with the increase over this same redshift range found by 
\citet{galametz09} and \citet{hart11}, although for a somewhat different 
range in luminosities. 

Two complications in the interpretation of the evolution of the AGN fraction 
are the extent to which the high-redshift clusters are equivalent 
to the progenitors of the low-redshift clusters and the incompleteness and 
evolution of the cluster galaxies. 
If the AGN fraction is a strong function of cluster mass, in addition to 
redshift, then mass dependence could manifest as redshift dependence. 
Based on previous analysis of the $1 < z < 1.5$ sample 
\citep{brodwin07,brodwin11,jee11}, these clusters are expected to be the 
high-redshift progenitors of present-day $10^{15}$ \msun\ clusters such as 
Coma. The cluster samples in the range 
$0 < z < 1$ studied by \citet{martini09} are similarly massive and 
consistent with the same population of clusters in the local universe. 
Similarity to Coma was explicitly used by \citet{hart09} to select their 
sample, which overlaps with the other low-redshift cluster samples. To 
the extent that there is a bias in these data, the higher-redshift clusters 
tend to be slightly more massive. As there is an anticorrelation between 
luminous AGN and environment in the local universe \citep{kauffmann04}, 
this anticorrelation would produce an underestimate of the true rate 
of evolution. 

The other complication originates in the comparison of the cluster galaxy 
populations across different surveys and from the local universe to $z=1.5$. 
In this study we have focused on the AGN fraction for cluster galaxies 
more luminous than one magnitude below the knee of the luminosity function 
at the cluster redshift. This choice was motivated by the previous study by 
\citet{martini09}, who adopted the assumption that $M_R^*(z) = M_R^*(0) - z$. 
\citet{haines12} adopted a similar threshold for the galaxy population, 
although their galaxy luminosity threshold was 1.5 magnitudes below the 
knee of the luminosity function (defined in the $K-$band: $M_K^*+1.5$). We 
estimate that the half magnitude difference for the galaxy population is a 
minor effect compared to the Poisson uncertainties, as the $z<0.5$ AGN 
fractions measured by these two studies are consistent. At intermediate 
redshifts $0.5 < z < 1$ the cluster spectroscopic data are incomplete for 
galaxies (although not for the much smaller number of X-ray sources), and 
there is thus a substantial correction to estimate the total galaxy population 
of some clusters. \citet{martini09} investigated this effect and estimated that 
it introduced on order a factor of two uncertainty into the AGN fraction 
estimate at $z>0.5$, which is much smaller than the observed 
evolution. While the present sample at $1 < z < 1.5$ similarly suffers from 
spectroscopic incompleteness, the high-quality photometric redshifts, 
substantial sample of field galaxies, and availability of some spectroscopic 
follow-up substantially mitigate this uncertainty such that the cluster 
galaxy population is probably more reliably known for the present sample 
than for the sample at $0.5 < z < 1$. 

\section{Discussion} \label{sec:dis} 

\begin{figure*}[ht] 
\epsscale{1.2}
\plotone{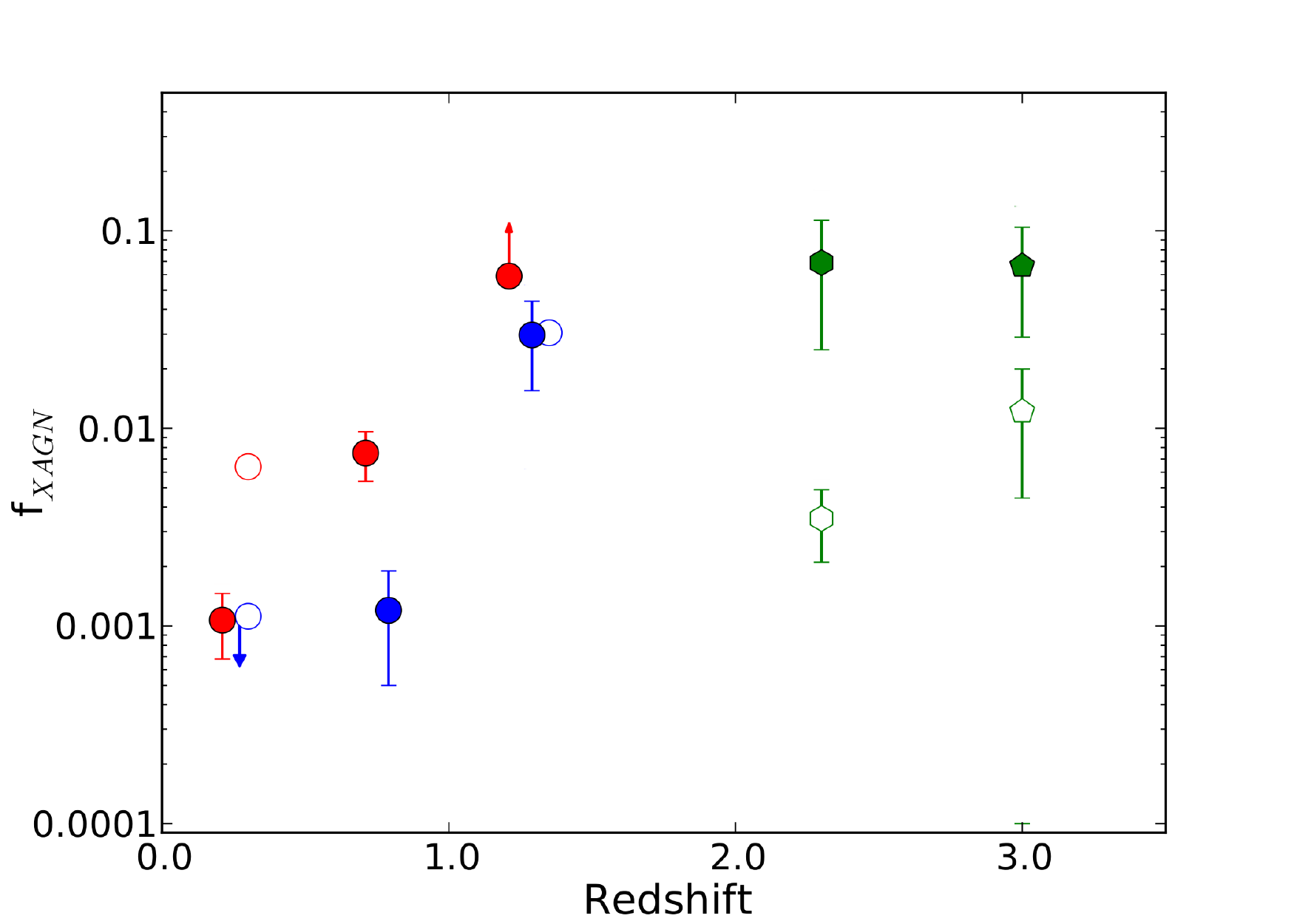}
\caption{Evolution of the X-ray AGN fraction in clusters ({\it solid symbols})
and the field ({\it open symbols}) from $z=0$ to $z\sim3$. All of the cluster 
measurements have been reproduced from Figure~\ref{fig:agnfracz}. 
The field AGN fractions at $z\sim0.3$ for 
$L_{X,H} \geq 10^{43}$ erg/s ({\it open, red circles}) and 
$L_{X,H} \geq 10^{44}$ erg/s ({\it open, blue circles}) 
are from D. Haggard \citep[{\it private communication}, see also][]{haggard10}. 
The formal uncertainties on these two field fractions are smaller than the 
size of the circles. 
The field AGN point at $z\sim1.25$ ({\it open blue circle}) is 
scaled from the cluster point by the ratio of the field and cluster 
fractions shown in Figure~\ref{fig:relfrac}. 
The $z=2.3$ protocluster and neighboring field fractions from 
\citet{digbynorth10} 
({\it filled and open green hexagons, respectively})
and the $z=3.09$ protocluster and neighboring field fractions from 
\citet{lehmer09} 
({\it filled and open green pentagons, respectively})
are also shown. These points are described further in \S\ref{sec:dis}. 
\label{fig:envz} 
}
\end{figure*}

The field X-ray AGN fraction also increases over the same redshift range 
of $0 < z < 1.5$ where the cluster AGN fraction has increased by factors 
of at least $30 - 45$ \citep[e.g.][]{ueda03}. At $z\sim1.25$, we showed in 
\S\ref{sec:rad} and Figure~\ref{fig:relfrac} that the cluster and field AGN 
fractions were comparable and both $\sim 3-4$\% for AGN with hard X-ray 
luminosities greater than a few times $10^{43}$ erg/s. Because our field 
measurement is from the immediate vicinity of massive clusters, some of the 
field galaxies may have already been processed through the cluster 
\citep[e.g.][]{diaferio01,patel09,balogh09,bahe12} and the field sample may 
not be representative of the true field population. Another estimate of the 
field AGN fraction at these 
redshifts was presented by \citet{bundy08} from an analysis of galaxies and 
AGN in the DEEP2 survey. They found that the AGN fraction is $\sim 1-3$\% for 
AGN with $L_{X,H}>10^{43}$ erg/s in host galaxies with stellar masses of 
$M_* \sim 10^{11.5}$ at $1 < z < 1.4$. This fraction 
appears consistent with our measurement for the field fraction around 
the clusters, although our data do not extend to $L_{X,H}=10^{43}$ erg/s 
for most clusters and the stellar mass range quoted by \citet{bundy08} is not 
an exact match to our luminosity threshold at $3.6\mu$m. 


In the local universe, very luminous AGN are rarely found in the field and 
very rarely found in clusters. 
An SDSS study by \citet{kauffmann04} found that AGN with 
$L$[\ion{O}{3}]$ > 10^7$ L$_\odot$ were approximately three times rarer in 
dense environments relative to less dense environments, where this 
luminosity threshold is approximately comparable to an AGN with 
$L_{X,H} = 10^{43}$ erg/s. 
A factor of three to four 
decrease in the AGN fraction with the same luminosity threshold was also found 
by \citet{best05b} over an order of magnitude change in local density with 
data from SDSS. However, both studies also find little environmental
dependence in the AGN fraction when only lower-luminosity AGN were examined. 
Note that for both of these studies, the high density regions are still 
not as dense as the centers of rich clusters of galaxies, which are 
embedded in very rare and very massive dark matter halos (on order 
$10^{15}$ \msun) in the local universe. 

The local field X-ray AGN fraction was measured by \citet{haggard10} based 
on data from the \chandra\ Multiwavelength Project \citep[ChaMP;][]{green04} 
and SDSS, although they do not use hard X-ray 
luminosities and probe a somewhat different range in galaxy luminosity and 
redshift. \citet{haggard10} compared their results to previous cluster 
studies \citep{martini06,martini07} and found the field and cluster AGN 
fractions are the same for low-luminosity AGN ($L_X > 10^{41-42}$ erg/s). Other 
studies have reached similar conclusions, namely the fraction of low-luminosity 
AGN is comparable in clusters, groups, and the field 
in the local universe \citep{sivakoff08,arnold09,miller12}. The sample studied 
by \citet{haggard10} also includes some higher-luminosity AGN that are more 
directly comparable to the AGN considered here. D. Haggard ({\it private 
communication}) has computed the field AGN fraction for 
similar X-ray and galaxy luminosity limits for $0 < z < 0.6$ 
and found $f_A (L_{X,H}>10^{43}) = 0.0064^{+0.0004}_{-0.0005}$ and 
$f_A (L_{X,H}>10^{44}) = 0.0011^{+0.0002}_{-0.0002}$. 
These points are shown in Figure~\ref{fig:envz} (note the formal uncertainties 
are smaller than the points) and demonstrate that the 
$L_{X,H}>10^{43}$ \ergs\ field AGN fraction is six times higher than the 
cluster value and the 
$L_{X,H}>10^{44}$ \ergs\ field AGN fraction is consistent with the $3\sigma$ 
upper limit for the cluster AGN fraction. 
While the two field fractions at $z \sim 0.3$ 
are calculated with the ChaMP survey's definition of the hard band of 2-8 keV, 
rather than the 2-10 keV band adopted throughout the rest of this paper, this 
is a very minor difference. 
The relative field and cluster AGN fractions in the local universe and at 
$z \sim 1.25$ show that while luminous AGN are anticorrelated with local density in the local universe, this is no longer the case at $z\sim1.25$. The masses 
of these high-redshift clusters are also in the range expected for the 
progenitors of the local clusters. 


Studies at even higher redshift support this trend and suggest that the 
present-day anticorrelation has reversed by $z>2$. \chandra\ observations of 
three protoclusters at $z>2$ have revealed luminous 
AGN associated with PKS 1138-262 at $z=2.16$ \citep{pentericci02,croft05}, 
the $z=2.3$ protocluster in the field of QSO HS 1700+643 \citep{digbynorth10}, 
and the $z=3.09$ SSA22 protocluster \citep{lehmer09}. 
\citet{pentericci02} compared the number of sources toward PKS 1138-262 
and found an excess of $\sim 50$\% compared to expectations from the 
AGN space density at this redshift. \citet{lehmer09} and \citet{digbynorth10} 
both measure the AGN fractions of the protoclusters and in field samples at 
the same redshift. 
\citet{lehmer09} detected X-ray emission from six LBGs and five LAEs 
(ten unique sources) toward the SSA22 protocluster at $z=3.09$ with a 
400ks \chandra\ observation. These sources 
have X-ray luminosities of $3-50 \times 10^{43}$ erg/s in the rest-frame 
8-32 keV band. They measure AGN fractions of $9.5^{+12.7}_{-6.1}$\% and 
$5.1^{+6.8}_{-3.3}$\% in LBGs and LAEs, respectively. These AGN fractions are 
larger by a factor of $6.1^{+10.3}_{-3.6}$ compared to the lower-density field 
at the same redshift. 
\citet{digbynorth10} studied the $z=2.30$ protocluster in the field of 
QSO HS 1700+643 and found five protocluster AGN in their $\sim 200$ks 
\chandra\ observation. They identified AGN with 
$L_{X,H} \geq 4.6\times10^{43}$ erg/s in a sample of members selected via 
the BX/MD method and as LAEs. The X-ray AGN fractions in the BX/MD and LAE 
samples are $6.9^{+9.2}_{-4.4}$ and $2.9^{+2.9}_{-1.6}$ \%, respectively, 
which are greater than the field fractions in similar galaxies at this 
redshift, particularly for the BX/MD sample. The protocluster and field 
fractions measured by \citet{lehmer09} and 
\citet[][for BX/MD galaxies]{digbynorth10} are shown on Figure~\ref{fig:envz}. 
While the field fractions for these high-redshift fields are measured in their 
immediate vicinity, this should be less important than for our sample because
both less time has elapsed for environmental pre-processing and the 
environments are not as evolved. 

These data on the evolution of the AGN fraction in the field, clusters, 
and protoclusters, while heterogeneous with respect to X-ray sensitivity 
thresholds, cluster selection, and host galaxy properties, broadly indicate 
that luminous AGN are 
anticorrelated with density in the local universe, found in cluster and field 
galaxies with approximately equal frequency at $1 < z < 1.5$, and are 
correlated with local density by $z > 2$. This relative evolution of the 
fraction of luminous AGN in field and cluster galaxies is consistent with the 
now-conventional picture that the most luminous AGN are fueled by gas-rich 
galaxy mergers \citep{sanders88} and the 
steady decline in the cold gas content of galaxies from high redshift to 
the present day. At the highest redshifts discussed here, which correspond to 
the redshifts of the protoclusters at$2 < z < 3.1$, a larger fraction 
of the baryonic mass fraction of massive galaxies is in molecular gas 
compared to their local analogs \citep{tacconi10}. 
The overdense protocluster environment leads to greater likelihood of 
interactions and mergers of these gas-rich galaxies and thus the fueling of 
luminous AGN. By $1 < z < 1.5$, the most overdense cluster 
environments have grown substantially and are large enough to have substantial 
hot gas reservoirs, as illustrated by the detection of extended X-ray emission 
from some clusters at $z>1$ \citep{gobat11,brodwin11,stanford12}. While 
there is still substantial star formation in many cluster members, including 
clusters in the present sample (Brodwin et al. 2013, {\it in preparation}), 
many other cluster galaxies have stopped substantial star formation, as 
indicated by the presence of a color-magnitude relation \citep{mei09}. 
This transition is likely because they have largely exhausted their cold gas 
supply. The increase in the cross section for galaxy interactions and mergers 
in the cluster environment is consequently counterbalanced by the smaller 
fraction of gas-rich galaxies, except perhaps on the outskirts of clusters 
\citep{wagg12}. Even at redshift $0.9 < z < 1.6$ there is some evidence for 
an enhancement of AGN in the infall region \citep{fassbender12}. 
By the present day, cluster galaxies have much less cold gas 
compared to field galaxies \citep{giovanardi83,oosterloo10} and the relative 
velocities of their member galaxies are too great to produce bound pairs 
that eventually merge. Aside from the central cluster galaxy, which is likely 
fueled by gas cooling from the ICM, luminous AGN may only be found in 
present-day clusters due to the infall of gas-rich field galaxies 
\citep{haines12}. This result is similar to the properties of star forming 
galaxies in clusters \citep{dressler99}. 

\section{Summary} 

We have investigated the X-ray and MIR-selected AGN population in a sample of 
13 clusters at $1 < z < 1.5$ identified by the \spitzer/IRAC Shallow Cluster 
Survey of \citet{eisenhardt08}. 
We find a total of 11 X-ray counterparts to cluster members, eight of which 
have spectroscopic redshifts. There are also 27 MIR AGN associated with cluster 
members, 12 of which have spectroscopic redshifts. All but three of the 
X-ray AGN are also MIR AGN. 

The X-ray AGN are quite luminous. All of the X-ray AGN have rest-frame, 
hard X-ray luminosities of $L_{X,H} > 10^{43}$ erg/s and four have 
$L_{X,H} > 10^{44}$ erg/s. AGN at these luminosities are extremely rare 
in low-redshift clusters, and in fact none have been found above 
$L_{X,H} > 10^{44}$ erg/s in a combined sample of 44 clusters at $z<0.5$ 
that we constructed from studies by \citet{martini09} and \citet{haines12}. 
These new observations demonstrate that the order of magnitude increase in the 
cluster AGN fraction from the present to $z\sim1$ continues to $z\sim1.5$. 

We have used photometric redshift estimates for galaxies out to five times 
the expected $r_{200}$ radius of the clusters, or $R = 2' - 10'$ from the 
cluster centers, to characterize the field population and likely 
field contamination due to the use of photometric redshifts to estimate 
cluster membership. These observations clearly indicate a substantial 
excess of galaxies, X-ray AGN, and MIR AGN associated with the clusters. 
We calculate the X-ray and MIR AGN fractions of all galaxies brighter than 
$M_{3.6}^*+1$, where $M_{3.6}^*$ corresponds to the knee of the luminosity 
function at the cluster redshift. We find that the field AGN fractions, 
defined from the sample of galaxies at the cluster redshift, but from 
$R = 2' - 6'$ (X-ray AGN) or $R = 2' - 10'$ (IR AGN) from the cluster center, 
are consistent with the cluster AGN 
fractions within $R \leq 2'$. This stands in sharp contrast to estimates at 
low redshift, where the luminous X-ray AGN fraction is substantially lower
in clusters relative to the field \citep[e.g.][]{martini09,haggard10}. 

The order of magnitude evolution of the cluster AGN fraction from $z=0$ to 
$z \sim 1.25$ is greater than the rate of evolution of the field AGN fraction. 
While the luminous AGN fraction is approximately six times higher in the 
field than in clusters in the local universe, we find comparable fractions 
in the field and clusters at $z \sim 1.25$. Studies of two protoclusters and 
samples of field galaxies at even higher redshifts indicate that the luminous 
AGN fraction is higher in protoclusters than the field at $z>2$ 
\citep{lehmer09,digbynorth10}. Taken together, these studies and our own 
demonstrate that there is a reversal of the local anticorrelation between 
luminous AGN and local density at high redshift. 
The relative evolution of the AGN fraction in the field and clusters is 
strong evidence of environment-dependent AGN evolution. 

\acknowledgements 

We thank Daryl Haggard for calculating the field AGN fraction based on 
our AGN and galaxy luminosity thresholds. We also appreciate a thoughtful 
and helpful review from the referee. PM appreciates support from the 
sabbatical visitor program at the North American ALMA Science Center (NAASC) 
at NRAO and the hospitality of both the NAASC and the University of Virginia 
while this work was completed. 
The work of PRME and DS was carried out at the Jet Propulsion Laboratory, 
California Institute of Technology, under a contract with NASA.
Part of this work was performed under the auspices of the U.S. Department of 
Energy by Lawrence Livermore National Laboratory under Contract 
DE-AC52-07NA27344. Support for this work was provided by the National 
Aeronautics and Space Administration through Chandra Award Number GO9-0150A 
issued by the Chandra X-ray Observatory Center, which is operated by the 
Smithsonian Astrophysical Observatory for and on behalf of the National 
Aeronautics Space Administration under contract NAS8-03060.

{\it Facilities:} \facility{Spitzer}, \facility{CXO}


\end{document}

%% file: martini.tab1.tex
\begin{deluxetable*}{lllclrr}
\tablecolumns{7}
\tablewidth{7.0truein}
\tabletypesize{\scriptsize}
\tablecaption{Properties of the High-Redshift Clusters\label{tbl:sample}}
\tablehead{
\colhead{Cluster} &
\colhead{$\alpha$} &
\colhead{$\delta$} &
\colhead{$z$} &
\colhead{ObsID} &
\colhead{$t_{exp}$ [ks]} &
\colhead{$M_{3.6\mu m}^*$ [mag]} \\ 
\colhead{(1)} &
\colhead{(2)} &
\colhead{(3)} &
\colhead{(4)} &
\colhead{(5)} &
\colhead{(6)} &
\colhead{(7)} \\
}
\startdata
ISCSJ1429.2+3357 & 14:29:15.2 & 33:57:08.5 & 1.06 & 10450 & 23 &  17.20  \\
ISCSJ1432.4+3332 & 14:32:29.2 & 33:32:36.0 & 1.11 & 10452 & 34 &  17.25  \\
ISCSJ1426.1+3403 & 14:26:09.5 & 34:03:41.1 & 1.14 & 10451,7945,6995 & 11,41,10 &  17.30  \\
ISCSJ1426.5+3339 & 14:26:30.4 & 33:39:33.2 & 1.16 & 10453 & 35 &  17.30  \\
ISCSJ1434.5+3427 & 14:34:30.4 & 34:27:12.3 & 1.24 & 10455 & 34 &  17.50  \\
ISCSJ1429.3+3437 & 14:29:18.5 & 34:37:25.8 & 1.26 & 10454 & 30 &  17.50  \\
ISCSJ1432.6+3436 & 14:32:38.4 & 34:36:49.0 & 1.35 & 10456 & 32 &  17.60  \\
ISCSJ1425.3+3428 & 14:25:19.3 & 34:28:38.2 & 1.36 & 10458 & 36 &  17.60  \\
ISCSJ1433.8+3325 & 14:33:51.1 & 33:25:51.1 & 1.37 & 7946 & 40 &  17.60  \\
ISCSJ1434.7+3519 & 14:34:46.3 & 35:19:33.5 & 1.37 & 10459 & 32 &  17.60  \\
ISCSJ1432.3+3253 & 14:32:18.3 & 32:53:07.8 & 1.40 & 10457 & 34 &  17.65  \\
ISCSJ1438.1+3414 & 14:38:08.7 & 34:14:19.2 & 1.41 & 10461 & 101(S) &  17.65  \\
ISCSJ1432.4+3250 & 14:32:24.2 & 32:50:03.7 & 1.49 & 10457 & 34 &  17.75  \\
\enddata
\tablecomments{
Properties of the clusters in the sample. Columns are: (1) Cluster name; 
(2--3) Right Ascension and Declination
(J2000) of the cluster center; (4) Redshift; (5) \chandra\ ObsID of the dataset(s)
used in the analysis; (6) Total integration time of the \chandra\ data; and
(7) Vega magnitude at $3.6\mu$m of $L^*$ at the cluster redshift from
\citet{eisenhardt08}. Clusters were observed with the \chandra\ ACIS-I 
camera with the exception of ISCSJ1438.1+3414 (ACIS-S)
}
\end{deluxetable*}

%% file: martini.tab2.tex
\begin{deluxetable*}{llllclrr}
\tablecolumns{8}
\tablewidth{7.0truein}
\tabletypesize{\scriptsize}
\tablecaption{AGN in the High-Redshift Clusters\label{tbl:cagn}}
\tablehead{
\colhead{ID} &
\colhead{Cluster} &
\colhead{Redshift} &
\colhead{$\alpha$} &
\colhead{$\delta$} &
\colhead{$I$} &
\colhead{[$3.6$]} &
\colhead{X/IR} \\ 
\colhead{(1)} &
\colhead{(2)} &
\colhead{(3)} &
\colhead{(4)} &
\colhead{(5)} &
\colhead{(6)} &
\colhead{(7)} &
\colhead{(8)} \\
}
\startdata
  J142916.1+335537 & ISCSJ1429.2+3357 & 1.06$^s$ & 14:29:16.1 & +33:55:37.3 & 21.30 (0.10) & 15.62 (0.03) & Both \\
  J143227.2+333307 & ISCSJ1432.4+3332 & 1.15$^p$ & 14:32:27.2 & +33:33:07.5 & 21.40 (0.08) & 16.01 (0.03) & X-ray \\
  J142611.6+340226 & ISCSJ1426.1+3403 & 1.14$^s$ & 14:26:11.6 & +34:02:26.0 & 21.90 (0.12) & 16.95 (0.03) & IR \\
  J142607.6+340309 & ISCSJ1426.1+3403 & 1.12$^s$ & 14:26:07.6 & +34:03:09.3 & 21.66 (0.09) & 16.44 (0.03) & IR \\
  J142602.8+340405 & ISCSJ1426.1+3403 & 1.26$^p$ & 14:26:02.8 & +34:04:05.8 & 22.30 (0.18) & 16.99 (0.03) & IR \\
  J142610.2+340355 & ISCSJ1426.1+3403 & 1.10$^p$ & 14:26:10.2 & +34:03:55.6 & 22.33 (0.17) & 17.60 (0.03) & IR \\
  J142608.3+340430 & ISCSJ1426.1+3403 & 1.09$^p$ & 14:26:08.3 & +34:04:30.0 & 20.83 (0.04) & 16.12 (0.03) & IR \\
  J142627.5+333912 & ISCSJ1426.5+3339 & 1.28$^p$ & 14:26:27.5 & +33:39:12.8 & 22.80 (0.27) & 17.08 (0.03) & IR \\
  J143430.3+342712 & ISCSJ1434.5+3427 & 1.24$^s$ & 14:34:30.3 & +34:27:12.0 & 21.97 (0.14) & 16.28 (0.03) & X-ray \\
  J142915.2+343709 & ISCSJ1429.3+3437 & 1.27$^s$ & 14:29:15.2 & +34:37:09.2 & 21.94 (0.11) & 16.78 (0.03) & Both \\
  J143232.7+343538 & ISCSJ1432.6+3436 & 1.22$^p$ & 14:32:32.7 & +34:35:38.4 & 22.70 (0.28) & 17.36 (0.03) & IR \\
  J143237.8+343630 & ISCSJ1432.6+3436 & 1.31$^p$ & 14:32:37.8 & +34:36:30.9 & 22.46 (0.23) & 16.43 (0.03) & IR \\
  J143238.8+343647 & ISCSJ1432.6+3436 & 1.28$^p$ & 14:32:38.8 & +34:36:47.6 & 22.68 (0.26) & 17.20 (0.03) & IR \\
  J142512.9+342735 & ISCSJ1425.3+3428 & 1.21$^p$ & 14:25:12.9 & +34:27:35.2 & 21.80 (0.07) & 17.00 (0.03) & X-ray \\
  J142516.5+342755 & ISCSJ1425.3+3428 & 1.36$^s$ & 14:25:16.5 & +34:27:55.7 & 22.06 (0.12) & 17.39 (0.03) & Both \\
  J142520.3+342942 & ISCSJ1425.3+3428 & 1.36$^s$ & 14:25:20.3 & +34:29:42.8 & 21.52 (0.06) & 16.92 (0.03) & Both \\
  J142520.2+343014 & ISCSJ1425.3+3428 & 1.36$^p$ & 14:25:20.2 & +34:30:14.3 & 23.48 (0.29) & 17.31 (0.03) & IR \\
  J142519.8+343024 & ISCSJ1425.3+3428 & 1.17$^p$ & 14:25:19.8 & +34:30:24.8 & 22.23 (0.09) & 16.90 (0.03) & IR \\
  J143351.5+332645 & ISCSJ1433.8+3325 & 1.37$^s$ & 14:33:51.5 & +33:26:45.8 & 20.19 (0.03) & 16.12 (0.03) & Both \\
  J143445.7+351921 & ISCSJ1434.7+3519 & 1.37$^s$ & 14:34:45.7 & +35:19:21.7 & 22.69 (0.18) & 16.94 (0.03) & IR \\
  J143450.0+351958 & ISCSJ1434.7+3519 & 1.43$^p$ & 14:34:50.0 & +35:19:58.8 & 22.94 (0.28) & 17.34 (0.03) & IR \\
  J143216.4+325434 & ISCSJ1432.3+3253 & 1.39$^s$ & 14:32:16.4 & +32:54:34.1 & 20.43 (0.05) & 16.35 (0.03) & Both \\
  J143217.1+325235 & ISCSJ1432.3+3253 & 1.27$^p$ & 14:32:17.1 & +32:52:35.1 & 22.88 (0.28) & 17.49 (0.03) & IR \\
  J143218.1+325315 & ISCSJ1432.3+3253 & 1.40$^s$ & 14:32:18.1 & +32:53:15.9 & 21.85 (0.16) & 16.69 (0.03) & IR \\
  J143816.8+341440 & ISCSJ1438.1+3414 & 1.41$^s$ & 14:38:16.8 & +34:14:40.3 & 22.41 (0.16) & 17.05 (0.03) & Both \\
  J143817.4+341337 & ISCSJ1438.1+3414 & 1.46$^p$ & 14:38:17.4 & +34:13:37.6 & 22.52 (0.16) & 17.58 (0.03) & IR \\
  J143802.7+341548 & ISCSJ1438.1+3414 & 1.42$^p$ & 14:38:02.7 & +34:15:48.2 & 21.73 (0.23) & 16.61 (0.03) & IR \\
  J143217.1+325055 & ISCSJ1432.4+3250 & 1.58$^p$ & 14:32:17.1 & +32:50:55.1 & 22.60 (0.19) & 17.49 (0.03) & Both \\
  J143224.8+325005 & ISCSJ1432.4+3250 & 1.31$^p$ & 14:32:24.8 & +32:50:05.0 & 22.59 (0.14) & 17.08 (0.03) & IR \\
\enddata
\tablecomments{
AGN in the high-redshift clusters. Columns are: 
(1) AGN ID; 
(2) Cluster name; 
(3) AGN Redshift, either photometric (p) or spectroscopic (s); 
(4--5) Right Ascension and Declination (J2000); 
(6--7) $I$ and [$3.6$] (Vega) mag; and
(8) AGN selection via X-ray, IR, or Both criteria.
For AGN with photometric redshifts, the redshift listed in column 3 may
not agree with the cluster redshift because it corresponds to the
peak of the redshift probability distribution function, while membership
is based on the fraction of the integrated probability at the cluster
redshift. See \S\ref{sec:data} for further details.
}
\end{deluxetable*}

%% file: martini.tab3.tex
\begin{deluxetable*}{lllrrrrrr}[h]
\tablecolumns{9}
\tablewidth{7.0truein}
\tabletypesize{\scriptsize}
\tablecaption{X-ray Properties of the Cluster AGN\label{tbl:xagn}}
\tablehead{
\colhead{ID} &
\colhead{Cluster} &
\colhead{Redshift} &
\colhead{Counts} &
\colhead{Sig} &
\colhead{$F_{0.5-8}$} &
\colhead{$F_{2-8}$} &
\colhead{$L_{0.5-8}$} &
\colhead{$L_{2-10}$} \\
\colhead{(1)} &
\colhead{(2)} &
\colhead{(3)} &
\colhead{(4)} &
\colhead{(5)} &
\colhead{(6)} &
\colhead{(7)} &
\colhead{(8)} &
\colhead{(9)} \\
}
\startdata
  J142916.1+335537 & ISCSJ1429.2+3357 & 1.06$^s$ &   89 &   44.0 & 6.71 (0.70) & 4.04 (0.42) & 3.20 (0.33) & 2.44 (0.26) \\
  J143227.2+333307 & ISCSJ1432.4+3332 & 1.15$^p$ &    6 &    3.5 & 0.25 (0.09) & 0.15 (0.06) & 0.14 (0.05) & 0.11 (0.04) \\
  J143430.3+342712 & ISCSJ1434.5+3427 & 1.24$^s$ &   11 &    6.0 & 0.49 (0.13) & 0.29 (0.08) & 0.33 (0.09) & 0.26 (0.07) \\
  J142915.2+343709 & ISCSJ1429.3+3437 & 1.27$^s$ &   75 &   37.0 & 4.11 (0.47) & 2.48 (0.28) & 2.91 (0.33) & 2.28 (0.26) \\
  J142512.9+342735 & ISCSJ1425.3+3428 & 1.21$^p$ &    4 &    2.4 & 0.20 (0.08) & 0.12 (0.05) & 0.12 (0.05) & 0.10 (0.04) \\
  J142516.5+342755 & ISCSJ1425.3+3428 & 1.36$^s$ &   25 &   12.8 & 0.99 (0.18) & 0.59 (0.11) & 0.81 (0.15) & 0.65 (0.12) \\
  J142520.3+342942 & ISCSJ1425.3+3428 & 1.36$^s$ &   25 &   12.4 & 0.90 (0.17) & 0.54 (0.10) & 0.72 (0.17) & 0.58 (0.14) \\
  J143351.5+332645 & ISCSJ1433.8+3325 & 1.37$^s$ &   81 &   39.1 & 2.65 (0.28) & 1.60 (0.17) & 2.21 (0.24) & 1.76 (0.19) \\
  J143216.4+325434 & ISCSJ1432.3+3253 & 1.39$^s$ &   63 &   25.9 & 2.55 (0.31) & 1.54 (0.19) & 2.21 (0.27) & 1.75 (0.21) \\
  J143816.8+341440 & ISCSJ1438.1+3414 & 1.41$^s$ &   10 &    4.5 & 0.30 (0.05) & 0.18 (0.03) & 0.26 (0.04) & 0.21 (0.03) \\
  J143217.1+325055 & ISCSJ1432.4+3250 & 1.58$^p$ &    9 &    5.0 & 0.43 (0.14) & 0.26 (0.08) & 0.49 (0.15) & 0.40 (0.13) \\
\enddata
\tablecomments{
X-ray properties of the AGN in the high-redshift clusters. Columns are: 
(1) AGN ID; 
(2) Cluster name; 
(3) AGN Redshift, either photometric (p) or spectroscopic (s); 
(4) Counts; 
(5) Significance of the X-ray detection; 
(6--7) Unabsorbed flux in the observed-frame 0.5-8 keV and 2-8 keV bands 
in units of $10^{-14}$ erg cm$^{-2}$ s$^{-1}$; 
(8--9) Unabsorbed luminosity in the rest-frame 0.5-8 keV and 2-10 keV 
bands in units of $10^{44}$ erg s$^{-1}$.
}
\end{deluxetable*}

%% file: martini.bbl
\begin{thebibliography}{109}
\expandafter\ifx\csname natexlab\endcsname\relax\def\natexlab#1{#1}\fi

\bibitem[{{Aird} {et~al.}(2012){Aird}, {Coil}, {Moustakas}, {Blanton},
  {Burles}, {Cool}, {Eisenstein}, {Smith}, {Wong}, \& {Zhu}}]{aird12}
{Aird}, J., {Coil}, A.~L., {Moustakas}, J., {Blanton}, M.~R., {Burles}, S.~M.,
  {Cool}, R.~J., {Eisenstein}, D.~J., {Smith}, M.~S.~M., {Wong}, K.~C., \&
  {Zhu}, G. 2012, \apj, 746, 90

\bibitem[{{Alonso-Herrero} {et~al.}(2008){Alonso-Herrero},
  {P{\'e}rez-Gonz{\'a}lez}, {Rieke}, {Alexander}, {Rigby}, {Papovich},
  {Donley}, \& {Rigopoulou}}]{alonsoherrero08}
{Alonso-Herrero}, A., {P{\'e}rez-Gonz{\'a}lez}, P.~G., {Rieke}, G.~H.,
  {Alexander}, D.~M., {Rigby}, J.~R., {Papovich}, C., {Donley}, J.~L., \&
  {Rigopoulou}, D. 2008, \apj, 677, 127

\bibitem[{{Arnold} {et~al.}(2009){Arnold}, {Martini}, {Mulchaey}, {Berti}, \&
  {Jeltema}}]{arnold09}
{Arnold}, T.~J., {Martini}, P., {Mulchaey}, J.~S., {Berti}, A., \& {Jeltema},
  T.~E. 2009, \apj, 707, 1691

\bibitem[{{Ashby} {et~al.}(2009){Ashby}, {Stern}, {Brodwin}, {Griffith},
  {Eisenhardt}, {Koz{\l}owski}, {Kochanek}, {Bock}, {Borys}, {Brand}, {Brown},
  {Cool}, {Cooray}, {Croft}, {Dey}, {Eisenstein}, {Gonzalez}, {Gorjian},
  {Grogin}, {Ivison}, {Jacob}, {Jannuzi}, {Mainzer}, {Moustakas},
  {R{\"o}ttgering}, {Seymour}, {Smith}, {Stanford}, {Stauffer}, {Sullivan},
  {van Breugel}, {Willner}, \& {Wright}}]{ashby09}
{Ashby}, M.~L.~N., {Stern}, D., {Brodwin}, M., {Griffith}, R., {Eisenhardt},
  P., {Koz{\l}owski}, S., {Kochanek}, C.~S., {Bock}, J.~J., {Borys}, C.,
  {Brand}, K., {Brown}, M.~J.~I., {Cool}, R., {Cooray}, A., {Croft}, S., {Dey},
  A., {Eisenstein}, D., {Gonzalez}, A.~H., {Gorjian}, V., {Grogin}, N.~A.,
  {Ivison}, R.~J., {Jacob}, J., {Jannuzi}, B.~T., {Mainzer}, A., {Moustakas},
  L.~A., {R{\"o}ttgering}, H.~J.~A., {Seymour}, N., {Smith}, H.~A., {Stanford},
  S.~A., {Stauffer}, J.~R., {Sullivan}, I., {van Breugel}, W., {Willner},
  S.~P., \& {Wright}, E.~L. 2009, \apj, 701, 428

\bibitem[{{Assef} {et~al.}(2010){Assef}, {Kochanek}, {Brodwin}, {Cool},
  {Forman}, {Gonzalez}, {Hickox}, {Jones}, {Le Floc'h}, {Moustakas}, {Murray},
  \& {Stern}}]{assef10}
{Assef}, R.~J., {Kochanek}, C.~S., {Brodwin}, M., {Cool}, R., {Forman}, W.,
  {Gonzalez}, A.~H., {Hickox}, R.~C., {Jones}, C., {Le Floc'h}, E.,
  {Moustakas}, J., {Murray}, S.~S., \& {Stern}, D. 2010, \apj, 713, 970

\bibitem[{{Atlee} \& {Martini}(2012)}]{atlee12}
{Atlee}, D.~W. \& {Martini}, P. 2012, \apj, in press

\bibitem[{{Atlee} {et~al.}(2011){Atlee}, {Martini}, {Assef}, {Kelson}, \&
  {Mulchaey}}]{atlee11}
{Atlee}, D.~W., {Martini}, P., {Assef}, R.~J., {Kelson}, D.~D., \& {Mulchaey},
  J.~S. 2011, \apj, 729, 22

\bibitem[Bahe et al.(2012)]{bahe12} Bahe, Y.~M., McCarthy, 
I.~G., Balogh, M.~L., \& Font, A.~S.\ 2012, arXiv:1210.8407 

\bibitem[Balogh et al.(2009)]{balogh09} Balogh, M.~L., McGee, 
S.~L., Wilman, D., et al.\ 2009, \mnras, 398, 754 

\bibitem[{{Barger} {et~al.}(2005){Barger}, {Cowie}, {Mushotzky}, {Yang},
  {Wang}, {Steffen}, \& {Capak}}]{barger05}
{Barger}, A.~J., {Cowie}, L.~L., {Mushotzky}, R.~F., {Yang}, Y., {Wang}, W.-H.,
  {Steffen}, A.~T., \& {Capak}, P. 2005, \aj, 129, 578

\bibitem[{{Barnes} \& {Hernquist}(1991)}]{barnes91}
{Barnes}, J.~E. \& {Hernquist}, L.~E. 1991, \apjl, 370, L65

\bibitem[{{Best} {et~al.}(2005{\natexlab{a}}){Best}, {Kauffmann}, {Heckman},
  {Brinchmann}, {Charlot}, {Ivezi{\'c}}, \& {White}}]{best05a}
{Best}, P.~N., {Kauffmann}, G., {Heckman}, T.~M., {Brinchmann}, J., {Charlot},
  S., {Ivezi{\'c}}, {\v Z}., \& {White}, S.~D.~M. 2005{\natexlab{a}}, \mnras,
  362, 25

\bibitem[{{Best} {et~al.}(2005{\natexlab{b}}){Best}, {Kauffmann}, {Heckman}, \&
  {Ivezi{\'c}}}]{best05b}
{Best}, P.~N., {Kauffmann}, G., {Heckman}, T.~M., \& {Ivezi{\'c}}, {\v Z}.
  2005{\natexlab{b}}, \mnras, 362, 9

\bibitem[{{Boyle} {et~al.}(1998){Boyle}, {Georgantopoulos}, {Blair}, {Stewart},
  {Griffiths}, {Shanks}, {Gunn}, \& {Almaini}}]{boyle98}
{Boyle}, B.~J., {Georgantopoulos}, I., {Blair}, A.~J., {Stewart}, G.~C.,
  {Griffiths}, R.~E., {Shanks}, T., {Gunn}, K.~F., \& {Almaini}, O. 1998,
  \mnras, 296, 1

\bibitem[{{Brodwin} {et~al.}(2006){Brodwin}, {Brown}, {Ashby}, {Bian}, {Brand},
  {Dey}, {Eisenhardt}, {Eisenstein}, {Gonzalez}, {Huang}, {Jannuzi},
  {Kochanek}, {McKenzie}, {Murray}, {Pahre}, {Smith}, {Soifer}, {Stanford},
  {Stern}, \& {Elston}}]{brodwin06}
{Brodwin}, M., {Brown}, M.~J.~I., {Ashby}, M.~L.~N., {Bian}, C., {Brand}, K.,
  {Dey}, A., {Eisenhardt}, P.~R., {Eisenstein}, D.~J., {Gonzalez}, A.~H.,
  {Huang}, J., {Jannuzi}, B.~T., {Kochanek}, C.~S., {McKenzie}, E., {Murray},
  S.~S., {Pahre}, M.~A., {Smith}, H.~A., {Soifer}, B.~T., {Stanford}, S.~A.,
  {Stern}, D., \& {Elston}, R.~J. 2006, \apj, 651, 791

\bibitem[{{Brodwin} {et~al.}(2007){Brodwin}, {Gonzalez}, {Moustakas},
  {Eisenhardt}, {Stanford}, {Stern}, \& {Brown}}]{brodwin07}
{Brodwin}, M., {Gonzalez}, A.~H., {Moustakas}, L.~A., {Eisenhardt}, P.~R.,
  {Stanford}, S.~A., {Stern}, D., \& {Brown}, M.~J.~I. 2007, \apjl, 671, L93

\bibitem[{{Brodwin} {et~al.}(2011){Brodwin}, {Stern}, {Vikhlinin}, {Stanford},
  {Gonzalez}, {Eisenhardt}, {Ashby}, {Bautz}, {Dey}, {Forman}, {Gettings},
  {Hickox}, {Jannuzi}, {Jones}, {Mancone}, {Miller}, {Moustakas}, {Ruel},
  {Snyder}, \& {Zeimann}}]{brodwin11}
{Brodwin}, M., {Stern}, D., {Vikhlinin}, A., {Stanford}, S.~A., {Gonzalez},
  A.~H., {Eisenhardt}, P.~R., {Ashby}, M.~L.~N., {Bautz}, M., {Dey}, A.,
  {Forman}, W.~R., {Gettings}, D., {Hickox}, R.~C., {Jannuzi}, B.~T., {Jones},
  C., {Mancone}, C., {Miller}, E.~D., {Moustakas}, L.~A., {Ruel}, J., {Snyder},
  G., \& {Zeimann}, G. 2011, \apj, 732, 33

\bibitem[{{Broos} {et~al.}(2010){Broos}, {Townsley}, {Feigelson}, {Getman},
  {Bauer}, \& {Garmire}}]{broos10}
{Broos}, P.~S., {Townsley}, L.~K., {Feigelson}, E.~D., {Getman}, K.~V.,
  {Bauer}, F.~E., \& {Garmire}, G.~P. 2010, \apj, 714, 1582

\bibitem[{{Bundy} {et~al.}(2008){Bundy}, {Georgakakis}, {Nandra}, {Ellis},
  {Conselice}, {Laird}, {Coil}, {Cooper}, {Faber}, {Newman}, {Pierce},
  {Primack}, \& {Yan}}]{bundy08}
{Bundy}, K., {Georgakakis}, A., {Nandra}, K., {Ellis}, R.~S., {Conselice},
  C.~J., {Laird}, E., {Coil}, A., {Cooper}, M.~C., {Faber}, S.~M., {Newman},
  J.~A., {Pierce}, C.~M., {Primack}, J.~R., \& {Yan}, R. 2008, \apj, 681, 931

\bibitem[{{Butcher} \& {Oemler}(1978)}]{butcher78}
{Butcher}, H. \& {Oemler}, A. 1978, \apj, 219, 18

\bibitem[{{Cappelluti} {et~al.}(2005){Cappelluti}, {Cappi}, {Dadina},
  {Malaguti}, {Branchesi}, {D'Elia}, \& {Palumbo}}]{cappelluti05}
{Cappelluti}, N., {Cappi}, M., {Dadina}, M., {Malaguti}, G., {Branchesi}, M.,
  {D'Elia}, V., \& {Palumbo}, G.~G.~C. 2005, \aap, 430, 39

\bibitem[{{Carlberg} {et~al.}(1997){Carlberg}, {Yee}, \&
  {Ellingson}}]{carlberg97}
{Carlberg}, R.~G., {Yee}, H.~K.~C., \& {Ellingson}, E. 1997, \apj, 478, 462

\bibitem[{{Cowie} \& {Songaila}(1977)}]{cowie77}
{Cowie}, L.~L. \& {Songaila}, A. 1977, \nat, 266, 501

\bibitem[{{Croft} {et~al.}(2005){Croft}, {Kurk}, {van Breugel}, {Stanford}, {de
  Vries}, {Pentericci}, \& {R{\"o}ttgering}}]{croft05}
{Croft}, S., {Kurk}, J., {van Breugel}, W., {Stanford}, S.~A., {de Vries}, W.,
  {Pentericci}, L., \& {R{\"o}ttgering}, H. 2005, \aj, 130, 867

\bibitem[{{Demarco} {et~al.}(2005){Demarco}, {Rosati}, {Lidman}, {Homeier},
  {Scannapieco}, {Ben{\'{\i}}tez}, {Mainieri}, {Nonino}, {Girardi}, {Stanford},
  {Tozzi}, {Borgani}, {Silk}, {Squires}, \& {Broadhurst}}]{demarco05}
{Demarco}, R., {Rosati}, P., {Lidman}, C., {Homeier}, N.~L., {Scannapieco}, E.,
  {Ben{\'{\i}}tez}, N., {Mainieri}, V., {Nonino}, M., {Girardi}, M.,
  {Stanford}, S.~A., {Tozzi}, P., {Borgani}, S., {Silk}, J., {Squires}, G., \&
  {Broadhurst}, T.~J. 2005, \aap, 432, 381

\bibitem[Diaferio et al.(2001)]{diaferio01} Diaferio, A., 
Kauffmann, G., Balogh, M.~L., et al.\ 2001, \mnras, 323, 999 


\bibitem[{{Digby-North} {et~al.}(2010){Digby-North}, {Nandra}, {Laird},
  {Steidel}, {Georgakakis}, {Bogosavljevi{\'c}}, {Erb}, {Shapley}, {Reddy}, \&
  {Aird}}]{digbynorth10}
{Digby-North}, J.~A., {Nandra}, K., {Laird}, E.~S., {Steidel}, C.~C.,
  {Georgakakis}, A., {Bogosavljevi{\'c}}, M., {Erb}, D.~K., {Shapley}, A.~E.,
  {Reddy}, N.~A., \& {Aird}, J. 2010, \mnras, 407, 846

\bibitem[{{Donley} {et~al.}(2012){Donley}, {Koekemoer}, {Brusa}, {Capak},
  {Cardamone}, {Civano}, {Ilbert}, {Impey}, {Kartaltepe}, {Miyaji}, {Salvato},
  {Sanders}, {Trump}, \& {Zamorani}}]{donley12}
{Donley}, J.~L., {Koekemoer}, A.~M., {Brusa}, M., {Capak}, P., {Cardamone},
  C.~N., {Civano}, F., {Ilbert}, O., {Impey}, C.~D., {Kartaltepe}, J.~S.,
  {Miyaji}, T., {Salvato}, M., {Sanders}, D.~B., {Trump}, J.~R., \& {Zamorani},
  G. 2012, \apj, 748, 142

\bibitem[{{Dressler}(1980)}]{dressler80}
{Dressler}, A. 1980, \apj, 236, 351

\bibitem[{{Dressler} \& {Gunn}(1983)}]{dressler83}
{Dressler}, A. \& {Gunn}, J.~E. 1983, \apj, 270, 7

\bibitem[{{Dressler} {et~al.}(1999){Dressler}, {Smail}, {Poggianti}, {Butcher},
  {Couch}, {Ellis}, \& {Oemler}}]{dressler99}
{Dressler}, A., {Smail}, I., {Poggianti}, B.~M., {Butcher}, H., {Couch}, W.~J.,
  {Ellis}, R.~S., \& {Oemler}, A.~J. 1999, \apjs, 122, 51

\bibitem[{{Dressler} {et~al.}(1985){Dressler}, {Thompson}, \&
  {Shectman}}]{dressler85}
{Dressler}, A., {Thompson}, I.~B., \& {Shectman}, S.~A. 1985, \apj, 288, 481

\bibitem[{{Eastman} {et~al.}(2007){Eastman}, {Martini}, {Sivakoff}, {Kelson},
  {Mulchaey}, \& {Tran}}]{eastman07}
{Eastman}, J., {Martini}, P., {Sivakoff}, G., {Kelson}, D.~D., {Mulchaey},
  J.~S., \& {Tran}, K.-V. 2007, \apjl, 664, L9

\bibitem[{{Eckart} {et~al.}(2006){Eckart}, {Stern}, {Helfand}, {Harrison},
  {Mao}, \& {Yost}}]{eckart06}
{Eckart}, M.~E., {Stern}, D., {Helfand}, D.~J., {Harrison}, F.~A., {Mao},
  P.~H., \& {Yost}, S.~A. 2006, \apjs, 165, 19

\bibitem[{{Eisenhardt} {et~al.}(2004){Eisenhardt}, {Stern}, {Brodwin}, {Fazio},
  {Rieke}, {Rieke}, {Werner}, {Wright}, {Allen}, {Arendt}, {Ashby}, {Barmby},
  {Forrest}, {Hora}, {Huang}, {Huchra}, {Pahre}, {Pipher}, {Reach}, {Smith},
  {Stauffer}, {Wang}, {Willner}, {Brown}, {Dey}, {Jannuzi}, \&
  {Tiede}}]{eisenhardt04}
{Eisenhardt}, P.~R., {Stern}, D., {Brodwin}, M., {Fazio}, G.~G., {Rieke},
  G.~H., {Rieke}, M.~J., {Werner}, M.~W., {Wright}, E.~L., {Allen}, L.~E.,
  {Arendt}, R.~G., {Ashby}, M.~L.~N., {Barmby}, P., {Forrest}, W.~J., {Hora},
  J.~L., {Huang}, J.-S., {Huchra}, J., {Pahre}, M.~A., {Pipher}, J.~L.,
  {Reach}, W.~T., {Smith}, H.~A., {Stauffer}, J.~R., {Wang}, Z., {Willner},
  S.~P., {Brown}, M.~J.~I., {Dey}, A., {Jannuzi}, B.~T., \& {Tiede}, G.~P.
  2004, \apjs, 154, 48

\bibitem[{{Eisenhardt} {et~al.}(2008){Eisenhardt}, {Brodwin}, {Gonzalez},
  {Stanford}, {Stern}, {Barmby}, {Brown}, {Dawson}, {Dey}, {Doi}, {Galametz},
  {Jannuzi}, {Kochanek}, {Meyers}, {Morokuma}, \& {Moustakas}}]{eisenhardt08}
{Eisenhardt}, P.~R.~M., {Brodwin}, M., {Gonzalez}, A.~H., {Stanford}, S.~A.,
  {Stern}, D., {Barmby}, P., {Brown}, M.~J.~I., {Dawson}, K., {Dey}, A., {Doi},
  M., {Galametz}, A., {Jannuzi}, B.~T., {Kochanek}, C.~S., {Meyers}, J.,
  {Morokuma}, T., \& {Moustakas}, L.~A. 2008, \apj, 684, 905

\bibitem[{{Elbaz} {et~al.}(2007){Elbaz}, {Daddi}, {Le Borgne}, {Dickinson},
  {Alexander}, {Chary}, {Starck}, {Brandt}, {Kitzbichler}, {MacDonald},
  {Nonino}, {Popesso}, {Stern}, \& {Vanzella}}]{elbaz07}
{Elbaz}, D., {Daddi}, E., {Le Borgne}, D., {Dickinson}, M., {Alexander}, D.~M.,
  {Chary}, R.-R., {Starck}, J.-L., {Brandt}, W.~N., {Kitzbichler}, M.,
  {MacDonald}, E., {Nonino}, M., {Popesso}, P., {Stern}, D., \& {Vanzella}, E.
  2007, \aap, 468, 33

\bibitem[{{Elmegreen} {et~al.}(1998){Elmegreen}, {Elmegreen}, {Brinks}, {Yuan},
  {Kaufman}, {Klaric}, {Montenegro}, {Struck}, \& {Thomasson}}]{elmegreen98}
{Elmegreen}, B.~G., {Elmegreen}, D.~M., {Brinks}, E., {Yuan}, C., {Kaufman},
  M., {Klaric}, M., {Montenegro}, L., {Struck}, C., \& {Thomasson}, M. 1998,
  \apjl, 503, L119

\bibitem[{{Elston} {et~al.}(2006){Elston}, {Gonzalez}, {McKenzie}, {Brodwin},
  {Brown}, {Cardona}, {Dey}, {Dickinson}, {Eisenhardt}, {Jannuzi}, {Lin},
  {Mohr}, {Raines}, {Stanford}, \& {Stern}}]{elston06}
{Elston}, R.~J., {Gonzalez}, A.~H., {McKenzie}, E., {Brodwin}, M., {Brown},
  M.~J.~I., {Cardona}, G., {Dey}, A., {Dickinson}, M., {Eisenhardt}, P.~R.,
  {Jannuzi}, B.~T., {Lin}, Y., {Mohr}, J.~J., {Raines}, S.~N., {Stanford},
  S.~A., \& {Stern}, D. 2006, \apj, 639, 816

\bibitem[Fassbender et al.(2012)]{fassbender12} Fassbender, R., {\v 
S}uhada, R., \& Nastasi, A.\ 2012, Advances in Astronomy, 2012

\bibitem[{{Farouki} \& {Shapiro}(1981)}]{farouki81}
{Farouki}, R. \& {Shapiro}, S.~L. 1981, \apj, 243, 32

\bibitem[{{Ferrarese} \& {Merritt}(2000)}]{ferrarese00}
{Ferrarese}, L. \& {Merritt}, D. 2000, \apjl, 539, L9

\bibitem[{{Franceschini} {et~al.}(1999){Franceschini}, {Hasinger}, {Miyaji}, \&
  {Malquori}}]{franceschini99}
{Franceschini}, A., {Hasinger}, G., {Miyaji}, T., \& {Malquori}, D. 1999,
  \mnras, 310, L5

\bibitem[{{Fuentes-Williams} \& {Stocke}(1988)}]{fuenteswilliams88}
{Fuentes-Williams}, T. \& {Stocke}, J.~T. 1988, \aj, 96, 1235

\bibitem[{{Galametz} {et~al.}(2009){Galametz}, {Stern}, {Eisenhardt},
  {Brodwin}, {Brown}, {Dey}, {Gonzalez}, {Jannuzi}, {Moustakas}, \&
  {Stanford}}]{galametz09}
{Galametz}, A., {Stern}, D., {Eisenhardt}, P.~R.~M., {Brodwin}, M., {Brown},
  M.~J.~I., {Dey}, A., {Gonzalez}, A.~H., {Jannuzi}, B.~T., {Moustakas}, L.~A.,
  \& {Stanford}, S.~A. 2009, \apj, 694, 1309

\bibitem[{{Gebhardt} {et~al.}(2000){Gebhardt}, {Bender}, {Bower}, {Dressler},
  {Faber}, {Filippenko}, {Green}, {Grillmair}, {Ho}, {Kormendy}, {Lauer},
  {Magorrian}, {Pinkney}, {Richstone}, \& {Tremaine}}]{gebhardt00b}
{Gebhardt}, K., {Bender}, R., {Bower}, G., {Dressler}, A., {Faber}, S.~M.,
  {Filippenko}, A.~V., {Green}, R., {Grillmair}, C., {Ho}, L.~C., {Kormendy},
  J., {Lauer}, T.~R., {Magorrian}, J., {Pinkney}, J., {Richstone}, D., \&
  {Tremaine}, S. 2000, \apjl, 539, L13

\bibitem[{{Gehrels}(1986)}]{gehrels86}
{Gehrels}, N. 1986, \apj, 303, 336

\bibitem[{{Genzel} {et~al.}(2008){Genzel}, {Burkert}, {Bouch{\'e}}, {Cresci},
  {F{\"o}rster Schreiber}, {Shapley}, {Shapiro}, {Tacconi}, {Buschkamp},
  {Cimatti}, {Daddi}, {Davies}, {Eisenhauer}, {Erb}, {Genel}, {Gerhard},
  {Hicks}, {Lutz}, {Naab}, {Ott}, {Rabien}, {Renzini}, {Steidel}, {Sternberg},
  \& {Lilly}}]{genzel08}
{Genzel}, R., {Burkert}, A., {Bouch{\'e}}, N., {Cresci}, G., {F{\"o}rster
  Schreiber}, N.~M., {Shapley}, A., {Shapiro}, K., {Tacconi}, L.~J.,
  {Buschkamp}, P., {Cimatti}, A., {Daddi}, E., {Davies}, R., {Eisenhauer}, F.,
  {Erb}, D.~K., {Genel}, S., {Gerhard}, O., {Hicks}, E., {Lutz}, D., {Naab},
  T., {Ott}, T., {Rabien}, S., {Renzini}, A., {Steidel}, C.~C., {Sternberg},
  A., \& {Lilly}, S.~J. 2008, \apj, 687, 59

\bibitem[{{Gilmour} {et~al.}(2009){Gilmour}, {Best}, \& {Almaini}}]{gilmour09}
{Gilmour}, R., {Best}, P., \& {Almaini}, O. 2009, ArXiv e-prints

\bibitem[{{Giovanardi} {et~al.}(1983){Giovanardi}, {Krumm}, \&
  {Salpeter}}]{giovanardi83}
{Giovanardi}, C., {Krumm}, N., \& {Salpeter}, E.~E. 1983, \aj, 88, 1719

\bibitem[{{Giovanelli} \& {Haynes}(1985)}]{giovanelli85}
{Giovanelli}, R. \& {Haynes}, M.~P. 1985, \apj, 292, 404

\bibitem[{{Gisler}(1978)}]{gisler78}
{Gisler}, G.~R. 1978, \mnras, 183, 633

\bibitem[{{Gobat} {et~al.}(2011){Gobat}, {Daddi}, {Onodera}, {Finoguenov},
  {Renzini}, {Arimoto}, {Bouwens}, {Brusa}, {Chary}, {Cimatti}, {Dickinson},
  {Kong}, \& {Mignoli}}]{gobat11}
{Gobat}, R., {Daddi}, E., {Onodera}, M., {Finoguenov}, A., {Renzini}, A.,
  {Arimoto}, N., {Bouwens}, R., {Brusa}, M., {Chary}, R.-R., {Cimatti}, A.,
  {Dickinson}, M., {Kong}, X., \& {Mignoli}, M. 2011, \aap, 526, A133

\bibitem[{{Gralla} {et~al.}(2011){Gralla}, {Gladders}, {Yee}, \&
  {Barrientos}}]{gralla11}
{Gralla}, M.~B., {Gladders}, M.~D., {Yee}, H.~K.~C., \& {Barrientos}, L.~F.
  2011, \apj, 734, 103

\bibitem[{{Green} {et~al.}(2004){Green}, {Silverman}, {Cameron}, {Kim},
  {Wilkes}, {Barkhouse}, {LaCluyz{\' e}}, {Morris}, {Mossman}, {Ghosh},
  {Grimes}, {Jannuzi}, {Tananbaum}, {Aldcroft}, {Baldwin}, {Chaffee}, {Dey},
  {Dosaj}, {Evans}, {Fan}, {Foltz}, {Gaetz}, {Hooper}, {Kashyap}, {Mathur},
  {McGarry}, {Romero-Colmenero}, {Smith}, {Smith}, {Smith}, {Torres},
  {Vikhlinin}, \& {Wik}}]{green04}
{Green}, P.~J., {Silverman}, J.~D., {Cameron}, R.~A., {Kim}, D.-W., {Wilkes},
  B.~J., {Barkhouse}, W.~A., {LaCluyz{\' e}}, A., {Morris}, D., {Mossman}, A.,
  {Ghosh}, H., {Grimes}, J.~P., {Jannuzi}, B.~T., {Tananbaum}, H., {Aldcroft},
  T.~L., {Baldwin}, J.~A., {Chaffee}, F.~H., {Dey}, A., {Dosaj}, A., {Evans},
  N.~R., {Fan}, X., {Foltz}, C., {Gaetz}, T., {Hooper}, E.~J., {Kashyap},
  V.~L., {Mathur}, S., {McGarry}, M.~B., {Romero-Colmenero}, E., {Smith},
  M.~G., {Smith}, P.~S., {Smith}, R.~C., {Torres}, G., {Vikhlinin}, A., \&
  {Wik}, D.~R. 2004, \apjs, 150, 43

\bibitem[{{Gunn} \& {Gott}(1972)}]{gunn72}
{Gunn}, J.~E. \& {Gott}, J.~R.~I. 1972, \apj, 176, 1

\bibitem[{{Haggard} {et~al.}(2010){Haggard}, {Green}, {Anderson}, {Constantin},
  {Aldcroft}, {Kim}, \& {Barkhouse}}]{haggard10}
{Haggard}, D., {Green}, P.~J., {Anderson}, S.~F., {Constantin}, A., {Aldcroft},
  T.~L., {Kim}, D., \& {Barkhouse}, W.~A. 2010, \apj, 723, 1447

\bibitem[{{Haines} {et~al.}(2012){Haines}, {Pereira}, {Sanderson}, {Smith},
  {Egami}, {Babul}, {Edge}, {Finoguenov}, {Moran}, \& {Okabe}}]{haines12}
{Haines}, C.~P., {Pereira}, M.~J., {Sanderson}, A.~J.~R., {Smith}, G.~P.,
  {Egami}, E., {Babul}, A., {Edge}, A.~C., {Finoguenov}, A., {Moran}, S.~M., \&
  {Okabe}, N. 2012, ArXiv e-prints

\bibitem[{{Haines} {et~al.}(2009){Haines}, {Smith}, {Egami}, {Ellis}, {Moran},
  {Sanderson}, {Merluzzi}, {Busarello}, \& {Smith}}]{haines09}
{Haines}, C.~P., {Smith}, G.~P., {Egami}, E., {Ellis}, R.~S., {Moran}, S.~M.,
  {Sanderson}, A.~J.~R., {Merluzzi}, P., {Busarello}, G., \& {Smith}, R.~J.
  2009, ArXiv e-prints

\bibitem[{{Hart} {et~al.}(2011){Hart}, {Stocke}, {Evrard}, {Ellingson}, \&
  {Barkhouse}}]{hart11}
{Hart}, Q.~N., {Stocke}, J.~T., {Evrard}, A.~E., {Ellingson}, E.~E., \&
  {Barkhouse}, W.~A. 2011, \apj, 740, 59

\bibitem[{{Hart} {et~al.}(2009){Hart}, {Stocke}, \& {Hallman}}]{hart09}
{Hart}, Q.~N., {Stocke}, J.~T., \& {Hallman}, E.~J. 2009, \apj, 705, 854

\bibitem[{{Hasinger} {et~al.}(2005){Hasinger}, {Miyaji}, \&
  {Schmidt}}]{hasinger05}
{Hasinger}, G., {Miyaji}, T., \& {Schmidt}, M. 2005, \aap, 441, 417

\bibitem[{{Heckman} {et~al.}(2004){Heckman}, {Kauffmann}, {Brinchmann},
  {Charlot}, {Tremonti}, \& {White}}]{heckman04}
{Heckman}, T.~M., {Kauffmann}, G., {Brinchmann}, J., {Charlot}, S., {Tremonti},
  C., \& {White}, S.~D.~M. 2004, \apj, 613, 109

\bibitem[{{Hickox} {et~al.}(2009){Hickox}, {Jones}, {Forman}, {Murray},
  {Kochanek}, {Eisenstein}, {Jannuzi}, {Dey}, {Brown}, {Stern}, {Eisenhardt},
  {Gorjian}, {Brodwin}, {Narayan}, {Cool}, {Kenter}, {Caldwell}, \&
  {Anderson}}]{hickox09}
{Hickox}, R.~C., {Jones}, C., {Forman}, W.~R., {Murray}, S.~S., {Kochanek},
  C.~S., {Eisenstein}, D., {Jannuzi}, B.~T., {Dey}, A., {Brown}, M.~J.~I.,
  {Stern}, D., {Eisenhardt}, P.~R., {Gorjian}, V., {Brodwin}, M., {Narayan},
  R., {Cool}, R.~J., {Kenter}, A., {Caldwell}, N., \& {Anderson}, M.~E. 2009,
  \apj, 696, 891

\bibitem[{{Hilton} {et~al.}(2010){Hilton}, {Lloyd-Davies}, {Stanford}, {Stott},
  {Collins}, {Romer}, {Hosmer}, {Hoyle}, {Kay}, {Liddle}, {Mehrtens}, {Miller},
  {Sahl{\'e}n}, \& {Viana}}]{hilton10}
{Hilton}, M., {Lloyd-Davies}, E., {Stanford}, S.~A., {Stott}, J.~P., {Collins},
  C.~A., {Romer}, A.~K., {Hosmer}, M., {Hoyle}, B., {Kay}, S.~T., {Liddle},
  A.~R., {Mehrtens}, N., {Miller}, C.~J., {Sahl{\'e}n}, M., \& {Viana},
  P.~T.~P. 2010, \apj, 718, 133

\bibitem[{{Hopkins} {et~al.}(2006){Hopkins}, {Hernquist}, {Cox}, {Di Matteo},
  {Robertson}, \& {Springel}}]{hopkins06a}
{Hopkins}, P.~F., {Hernquist}, L., {Cox}, T.~J., {Di Matteo}, T., {Robertson},
  B., \& {Springel}, V. 2006, \apjs, 163, 1

\bibitem[{{Hopkins} \& {Quataert}(2011)}]{hopkins11}
{Hopkins}, P.~F. \& {Quataert}, E. 2011, \mnras, 415, 1027

\bibitem[{{Jannuzi} \& {Dey}(1999)}]{jannuzi99}
{Jannuzi}, B.~T. \& {Dey}, A. 1999, in Astronomical Society of the Pacific
  Conference Series, Vol. 191, Photometric Redshifts and the Detection of High
  Redshift Galaxies, ed. {R.~Weymann, L.~Storrie-Lombardi, M.~Sawicki, \&
  R.~Brunner}, 111--+

\bibitem[{{Jee} {et~al.}(2011){Jee}, {Dawson}, {Hoekstra}, {Perlmutter},
  {Rosati}, {Brodwin}, {Suzuki}, {Koester}, {Postman}, {Lubin}, {Meyers},
  {Stanford}, {Barbary}, {Barrientos}, {Eisenhardt}, {Ford}, {Gilbank},
  {Gladders}, {Gonzalez}, {Harris}, {Huang}, {Lidman}, {Rykoff}, {Rubin}, \&
  {Spadafora}}]{jee11}
{Jee}, M.~J., {Dawson}, K.~S., {Hoekstra}, H., {Perlmutter}, S., {Rosati}, P.,
  {Brodwin}, M., {Suzuki}, N., {Koester}, B., {Postman}, M., {Lubin}, L.,
  {Meyers}, J., {Stanford}, S.~A., {Barbary}, K., {Barrientos}, F.,
  {Eisenhardt}, P., {Ford}, H.~C., {Gilbank}, D.~G., {Gladders}, M.~D.,
  {Gonzalez}, A., {Harris}, D.~W., {Huang}, X., {Lidman}, C., {Rykoff}, E.~S.,
  {Rubin}, D., \& {Spadafora}, A.~L. 2011, \apj, 737, 59

\bibitem[{{Johnson} {et~al.}(2003){Johnson}, {Best}, \& {Almaini}}]{johnson03}
{Johnson}, O., {Best}, P.~N., \& {Almaini}, O. 2003, \mnras, 343, 924

\bibitem[{{Kauffmann} {et~al.}(2003){Kauffmann}, {Heckman}, {Tremonti},
  {Brinchmann}, {Charlot}, {White}, {Ridgway}, {Brinkmann}, {Fukugita}, {Hall},
  {Ivezi{\'c}}, {Richards}, \& {Schneider}}]{kauffmann03}
{Kauffmann}, G., {Heckman}, T.~M., {Tremonti}, C., {Brinchmann}, J., {Charlot},
  S., {White}, S.~D.~M., {Ridgway}, S.~E., {Brinkmann}, J., {Fukugita}, M.,
  {Hall}, P.~B., {Ivezi{\'c}}, {\v Z}., {Richards}, G.~T., \& {Schneider},
  D.~P. 2003, \mnras, 346, 1055

\bibitem[{{Kauffmann} {et~al.}(2004){Kauffmann}, {White}, {Heckman},
  {M{\'e}nard}, {Brinchmann}, {Charlot}, {Tremonti}, \&
  {Brinkmann}}]{kauffmann04}
{Kauffmann}, G., {White}, S.~D.~M., {Heckman}, T.~M., {M{\'e}nard}, B.,
  {Brinchmann}, J., {Charlot}, S., {Tremonti}, C., \& {Brinkmann}, J. 2004,
  \mnras, 353, 713

\bibitem[{{Kelson} {et~al.}(1997){Kelson}, {van Dokkum}, {Franx},
  {Illingworth}, \& {Fabricant}}]{kelson97}
{Kelson}, D.~D., {van Dokkum}, P.~G., {Franx}, M., {Illingworth}, G.~D., \&
  {Fabricant}, D. 1997, \apjl, 478, L13

\bibitem[{{Kenter} {et~al.}(2005){Kenter}, {Murray}, {Forman}, {Jones},
  {Green}, {Kochanek}, {Vikhlinin}, {Fabricant}, {Fazio}, {Brand}, {Brown},
  {Dey}, {Jannuzi}, {Najita}, {McNamara}, {Shields}, \& {Rieke}}]{kenter05}
{Kenter}, A., {Murray}, S.~S., {Forman}, W.~R., {Jones}, C., {Green}, P.,
  {Kochanek}, C.~S., {Vikhlinin}, A., {Fabricant}, D., {Fazio}, G., {Brand},
  K., {Brown}, M.~J.~I., {Dey}, A., {Jannuzi}, B.~T., {Najita}, J., {McNamara},
  B., {Shields}, J., \& {Rieke}, M. 2005, \apjs, 161, 9

\bibitem[{{Kochanek} {et~al.}(2012){Kochanek}, {Eisenstein}, {Cool},
  {Caldwell}, {Assef}, {Jannuzi}, {Jones}, {Murray}, {Forman}, {Dey}, {Brown},
  {Eisenhardt}, {Gonzalez}, {Green}, \& {Stern}}]{kochanek12}
{Kochanek}, C.~S., {Eisenstein}, D.~J., {Cool}, R.~J., {Caldwell}, N., {Assef},
  R.~J., {Jannuzi}, B.~T., {Jones}, C., {Murray}, S.~S., {Forman}, W.~R.,
  {Dey}, A., {Brown}, M.~J.~I., {Eisenhardt}, P., {Gonzalez}, A.~H., {Green},
  P., \& {Stern}, D. 2012, \apjs, 200, 8

\bibitem[{{Lacy} {et~al.}(2004){Lacy}, {Storrie-Lombardi}, {Sajina},
  {Appleton}, {Armus}, {Chapman}, {Choi}, {Fadda}, {Fang}, {Frayer},
  {Heinrichsen}, {Helou}, {Im}, {Marleau}, {Masci}, {Shupe}, {Soifer},
  {Surace}, {Teplitz}, {Wilson}, \& {Yan}}]{lacy04}
{Lacy}, M., {Storrie-Lombardi}, L.~J., {Sajina}, A., {Appleton}, P.~N.,
  {Armus}, L., {Chapman}, S.~C., {Choi}, P.~I., {Fadda}, D., {Fang}, F.,
  {Frayer}, D.~T., {Heinrichsen}, I., {Helou}, G., {Im}, M., {Marleau}, F.~R.,
  {Masci}, F., {Shupe}, D.~L., {Soifer}, B.~T., {Surace}, J., {Teplitz}, H.~I.,
  {Wilson}, G., \& {Yan}, L. 2004, \apjs, 154, 166

\bibitem[{{Larson} {et~al.}(1980){Larson}, {Tinsley}, \& {Caldwell}}]{larson80}
{Larson}, R.~B., {Tinsley}, B.~M., \& {Caldwell}, C.~N. 1980, \apj, 237, 692

\bibitem[{{Lehmer} {et~al.}(2009){Lehmer}, {Alexander}, {Geach}, {Smail},
  {Basu-Zych}, {Bauer}, {Chapman}, {Matsuda}, {Scharf}, {Volonteri}, \&
  {Yamada}}]{lehmer09}
{Lehmer}, B.~D., {Alexander}, D.~M., {Geach}, J.~E., {Smail}, I., {Basu-Zych},
  A., {Bauer}, F.~E., {Chapman}, S.~C., {Matsuda}, Y., {Scharf}, C.~A.,
  {Volonteri}, M., \& {Yamada}, T. 2009, \apj, 691, 687

\bibitem[{{Mancone} {et~al.}(2010){Mancone}, {Gonzalez}, {Brodwin}, {Stanford},
  {Eisenhardt}, {Stern}, \& {Jones}}]{mancone10}
{Mancone}, C.~L., {Gonzalez}, A.~H., {Brodwin}, M., {Stanford}, S.~A.,
  {Eisenhardt}, P.~R.~M., {Stern}, D., \& {Jones}, C. 2010, \apj, 720, 284

\bibitem[{{Martini}(2004)}]{martini04c}
{Martini}, P. 2004, in IAU Symposium, ed. T.~{Storchi-Bergmann}, L.~C. {Ho}, \&
  H.~R. {Schmitt}, 235--241

\bibitem[{{Martini} {et~al.}(2006){Martini}, {Kelson}, {Kim}, {Mulchaey}, \&
  {Athey}}]{martini06}
{Martini}, P., {Kelson}, D.~D., {Kim}, E., {Mulchaey}, J.~S., \& {Athey}, A.~A.
  2006, \apj, 644, 116

\bibitem[{{Martini} {et~al.}(2002){Martini}, {Kelson}, {Mulchaey}, \&
  {Trager}}]{martini02}
{Martini}, P., {Kelson}, D.~D., {Mulchaey}, J.~S., \& {Trager}, S.~C. 2002,
  \apjl, 576, L109

\bibitem[{{Martini} {et~al.}(2007){Martini}, {Mulchaey}, \&
  {Kelson}}]{martini07}
{Martini}, P., {Mulchaey}, J.~S., \& {Kelson}, D.~D. 2007, \apj, 664, 761

\bibitem[{{Martini} {et~al.}(2003){Martini}, {Regan}, {Mulchaey}, \&
  {Pogge}}]{martini03b}
{Martini}, P., {Regan}, M.~W., {Mulchaey}, J.~S., \& {Pogge}, R.~W. 2003, \apj,
  589, 774

\bibitem[{{Martini} {et~al.}(2009){Martini}, {Sivakoff}, \&
  {Mulchaey}}]{martini09}
{Martini}, P., {Sivakoff}, G.~R., \& {Mulchaey}, J.~S. 2009, \apj, 701, 66

\bibitem[{{Mei} {et~al.}(2009){Mei}, {Holden}, {Blakeslee}, {Ford}, {Franx},
  {Homeier}, {Illingworth}, {Jee}, {Overzier}, {Postman}, {Rosati}, {Van der
  Wel}, \& {Bartlett}}]{mei09}
{Mei}, S., {Holden}, B.~P., {Blakeslee}, J.~P., {Ford}, H.~C., {Franx}, M.,
  {Homeier}, N.~L., {Illingworth}, G.~D., {Jee}, M.~J., {Overzier}, R.,
  {Postman}, M., {Rosati}, P., {Van der Wel}, A., \& {Bartlett}, J.~G. 2009,
  \apj, 690, 42

\bibitem[{{Merloni} {et~al.}(2004){Merloni}, {Rudnick}, \& {Di
  Matteo}}]{merloni04}
{Merloni}, A., {Rudnick}, G., \& {Di Matteo}, T. 2004, \mnras, 354, L37

\bibitem[{{Merritt}(1983)}]{merritt83a}
{Merritt}, D. 1983, \apj, 264, 24

\bibitem[{{Miller} {et~al.}(2003){Miller}, {Nichol}, {G{\' o}mez}, {Hopkins},
  \& {Bernardi}}]{miller03a}
{Miller}, C.~J., {Nichol}, R.~C., {G{\' o}mez}, P.~L., {Hopkins}, A.~M., \&
  {Bernardi}, M. 2003, \apj, 597, 142

\bibitem[{{Miller} {et~al.}(2012){Miller}, {Bautz}, {Forman}, {Jones},
  {Benson}, {Marrone}, {Reichardt}, {High}, {Brodwin}, \&
  {Carlstrom}}]{miller12}
{Miller}, E.~D., {Bautz}, M., {Forman}, W., {Jones}, C., {Benson}, B.,
  {Marrone}, D., {Reichardt}, C., {High}, F.~W., {Brodwin}, M., \& {Carlstrom},
  J. 2012, in American Astronomical Society Meeting Abstracts, Vol. 220,
  American Astronomical Society Meeting Abstracts, \#435.04

\bibitem[{{Moore} {et~al.}(1996){Moore}, {Katz}, {Lake}, {Dressler}, \&
  {Oemler}}]{moore96}
{Moore}, B., {Katz}, N., {Lake}, G., {Dressler}, A., \& {Oemler}, Jr., A. 1996,
  \nat, 379, 613

\bibitem[{{Mulchaey} \& {Regan}(1997)}]{mulchaey97a}
{Mulchaey}, J.~S. \& {Regan}, M.~W. 1997, \apjl, 482, L135

\bibitem[{{Murray} {et~al.}(2005){Murray}, {Kenter}, {Forman}, {Jones},
  {Green}, {Kochanek}, {Vikhlinin}, {Fabricant}, {Fazio}, {Brand}, {Brown},
  {Dey}, {Jannuzi}, {Najita}, {McNamara}, {Shields}, \& {Rieke}}]{murray05}
{Murray}, S.~S., {Kenter}, A., {Forman}, W.~R., {Jones}, C., {Green}, P.~J.,
  {Kochanek}, C.~S., {Vikhlinin}, A., {Fabricant}, D., {Fazio}, G., {Brand},
  K., {Brown}, M.~J.~I., {Dey}, A., {Jannuzi}, B.~T., {Najita}, J., {McNamara},
  B., {Shields}, J., \& {Rieke}, M. 2005, \apjs, 161, 1

\bibitem[{{Oosterloo} {et~al.}(2010){Oosterloo}, {Morganti}, {Crocker},
  {J{\"u}tte}, {Cappellari}, {de Zeeuw}, {Krajnovi{\'c}}, {McDermid},
  {Kuntschner}, {Sarzi}, \& {Weijmans}}]{oosterloo10}
{Oosterloo}, T., {Morganti}, R., {Crocker}, A., {J{\"u}tte}, E., {Cappellari},
  M., {de Zeeuw}, T., {Krajnovi{\'c}}, D., {McDermid}, R., {Kuntschner}, H.,
  {Sarzi}, M., \& {Weijmans}, A.-M. 2010, \mnras, 409, 500

\bibitem[Patel et al.(2009)]{patel09} Patel, S.~G., Kelson, 
D.~D., Holden, B.~P., et al.\ 2009, \apj, 694, 1349 

\bibitem[{{Pentericci} {et~al.}(2002){Pentericci}, {Kurk}, {Carilli}, {Harris},
  {Miley}, \& {R{\"o}ttgering}}]{pentericci02}
{Pentericci}, L., {Kurk}, J.~D., {Carilli}, C.~L., {Harris}, D.~E., {Miley},
  G.~K., \& {R{\"o}ttgering}, H.~J.~A. 2002, \aap, 396, 109

\bibitem[{{Popesso} \& {Biviano}(2006)}]{popesso06}
{Popesso}, P. \& {Biviano}, A. 2006, \aap, 460, L23

\bibitem[{{Richstone}(1976)}]{richstone76}
{Richstone}, D.~O. 1976, \apj, 204, 642

\bibitem[Rumbaugh et al.(2012)]{rumbaugh12} Rumbaugh, N., 
Kocevski, D.~D., Gal, R.~R., et al.\ 2012, \apj, 746, 155 

\bibitem[{{Saintonge} {et~al.}(2008){Saintonge}, {Tran}, \&
  {Holden}}]{saintonge08}
{Saintonge}, A., {Tran}, K.-V.~H., \& {Holden}, B.~P. 2008, \apjl, 685, L113

\bibitem[{{Sanders} {et~al.}(1988){Sanders}, {Soifer}, {Elias}, {Madore},
  {Matthews}, {Neugebauer}, \& {Scoville}}]{sanders88}
{Sanders}, D.~B., {Soifer}, B.~T., {Elias}, J.~H., {Madore}, B.~F., {Matthews},
  K., {Neugebauer}, G., \& {Scoville}, N.~Z. 1988, \apj, 325, 74

\bibitem[{{Silverman} {et~al.}(2008){Silverman}, {Green}, {Barkhouse}, {Kim},
  {Kim}, {Wilkes}, {Cameron}, {Hasinger}, {Jannuzi}, {Smith}, {Smith}, \&
  {Tananbaum}}]{silverman08b}
{Silverman}, J.~D., {Green}, P.~J., {Barkhouse}, W.~A., {Kim}, D.-W., {Kim},
  M., {Wilkes}, B.~J., {Cameron}, R.~A., {Hasinger}, G., {Jannuzi}, B.~T.,
  {Smith}, M.~G., {Smith}, P.~S., \& {Tananbaum}, H. 2008, \apj, 679, 118

\bibitem[{{Simkin} {et~al.}(1980){Simkin}, {Su}, \& {Schwarz}}]{simkin80}
{Simkin}, S.~M., {Su}, H.~J., \& {Schwarz}, M.~P. 1980, \apj, 237, 404

\bibitem[{{Sivakoff} {et~al.}(2008){Sivakoff}, {Martini}, {Zabludoff},
  {Kelson}, \& {Mulchaey}}]{sivakoff08}
{Sivakoff}, G.~R., {Martini}, P., {Zabludoff}, A.~I., {Kelson}, D.~D., \&
  {Mulchaey}, J.~S. 2008, \apj, 682, 803

\bibitem[{{Stanford} {et~al.}(2012){Stanford}, {Brodwin}, {Gonzalez},
  {Zeimann}, {Stern}, {Dey}, {Eisenhardt}, {Snyder}, \& {Mancone}}]{stanford12}
{Stanford}, S.~A., {Brodwin}, M., {Gonzalez}, A.~H., {Zeimann}, G., {Stern},
  D., {Dey}, A., {Eisenhardt}, P.~R., {Snyder}, G.~F., \& {Mancone}, C. 2012,
  \apj, 753, 164

\bibitem[{{Stanford} {et~al.}(2005){Stanford}, {Eisenhardt}, {Brodwin},
  {Gonzalez}, {Stern}, {Jannuzi}, {Dey}, {Brown}, {McKenzie}, \&
  {Elston}}]{stanford05}
{Stanford}, S.~A., {Eisenhardt}, P.~R., {Brodwin}, M., {Gonzalez}, A.~H.,
  {Stern}, D., {Jannuzi}, B.~T., {Dey}, A., {Brown}, M.~J.~I., {McKenzie}, E.,
  \& {Elston}, R. 2005, \apjl, 634, L129

\bibitem[{{Stern} {et~al.}(2005){Stern}, {Eisenhardt}, {Gorjian}, {Kochanek},
  {Caldwell}, {Eisenstein}, {Brodwin}, {Brown}, {Cool}, {Dey}, {Green},
  {Jannuzi}, {Murray}, {Pahre}, \& {Willner}}]{stern05}
{Stern}, D., {Eisenhardt}, P., {Gorjian}, V., {Kochanek}, C.~S., {Caldwell},
  N., {Eisenstein}, D., {Brodwin}, M., {Brown}, M.~J.~I., {Cool}, R., {Dey},
  A., {Green}, P., {Jannuzi}, B.~T., {Murray}, S.~S., {Pahre}, M.~A., \&
  {Willner}, S.~P. 2005, \apj, 631, 163

\bibitem[{{Tacconi} {et~al.}(2010){Tacconi}, {Genzel}, {Neri}, {Cox}, {Cooper},
  {Shapiro}, {Bolatto}, {Bouch{\'e}}, {Bournaud}, {Burkert}, {Combes},
  {Comerford}, {Davis}, {Schreiber}, {Garcia-Burillo}, {Gracia-Carpio}, {Lutz},
  {Naab}, {Omont}, {Shapley}, {Sternberg}, \& {Weiner}}]{tacconi10}
{Tacconi}, L.~J., {Genzel}, R., {Neri}, R., {Cox}, P., {Cooper}, M.~C.,
  {Shapiro}, K., {Bolatto}, A., {Bouch{\'e}}, N., {Bournaud}, F., {Burkert},
  A., {Combes}, F., {Comerford}, J., {Davis}, M., {Schreiber}, N.~M.~F.,
  {Garcia-Burillo}, S., {Gracia-Carpio}, J., {Lutz}, D., {Naab}, T., {Omont},
  A., {Shapley}, A., {Sternberg}, A., \& {Weiner}, B. 2010, \nat, 463, 781

\bibitem[Tanaka et al.(2012)]{tanaka12} Tanaka, M., Finoguenov, 
A., Mirkazemi, M., et al.\ 2012, arXiv:1210.0302 

\bibitem[{{Terlevich} {et~al.}(1990){Terlevich}, {Diaz}, \&
  {Terlevich}}]{terlevich90}
{Terlevich}, E., {Diaz}, A.~I., \& {Terlevich}, R. 1990, \mnras, 242, 271

\bibitem[{{Tran} {et~al.}(2010){Tran}, {Papovich}, {Saintonge}, {Brodwin},
  {Dunlop}, {Farrah}, {Finkelstein}, {Finkelstein}, {Lotz}, {McLure},
  {Momcheva}, \& {Willmer}}]{tran10}
{Tran}, K.-V.~H., {Papovich}, C., {Saintonge}, A., {Brodwin}, M., {Dunlop},
  J.~S., {Farrah}, D., {Finkelstein}, K.~D., {Finkelstein}, S.~L., {Lotz}, J.,
  {McLure}, R.~J., {Momcheva}, I., \& {Willmer}, C.~N.~A. 2010, \apjl, 719,
  L126

\bibitem[{{Tremaine} {et~al.}(2002){Tremaine}, {Gebhardt}, {Bender}, {Bower},
  {Dressler}, {Faber}, {Filippenko}, {Green}, {Grillmair}, {Ho}, {Kormendy},
  {Lauer}, {Magorrian}, {Pinkney}, \& {Richstone}}]{tremaine02}
{Tremaine}, S., {Gebhardt}, K., {Bender}, R., {Bower}, G., {Dressler}, A.,
  {Faber}, S.~M., {Filippenko}, A.~V., {Green}, R., {Grillmair}, C., {Ho},
  L.~C., {Kormendy}, J., {Lauer}, T.~R., {Magorrian}, J., {Pinkney}, J., \&
  {Richstone}, D. 2002, \apj, 574, 740

\bibitem[{{Ueda} {et~al.}(2003){Ueda}, {Akiyama}, {Ohta}, \& {Miyaji}}]{ueda03}
{Ueda}, Y., {Akiyama}, M., {Ohta}, K., \& {Miyaji}, T. 2003, \apj, 598, 886

\bibitem[{{van Dokkum} \& {Franx}(1996)}]{vandokkum96}
{van Dokkum}, P.~G. \& {Franx}, M. 1996, \mnras, 281, 985

\bibitem[{{Veilleux} {et~al.}(2009){Veilleux}, {Rupke}, {Kim}, {Genzel},
  {Sturm}, {Lutz}, {Contursi}, {Schweitzer}, {Tacconi}, {Netzer}, {Sternberg},
  {Mihos}, {Baker}, {Mazzarella}, {Lord}, {Sanders}, {Stockton}, {Joseph}, \&
  {Barnes}}]{veilleux09}
{Veilleux}, S., {Rupke}, D.~S.~N., {Kim}, D.-C., {Genzel}, R., {Sturm}, E.,
  {Lutz}, D., {Contursi}, A., {Schweitzer}, M., {Tacconi}, L.~J., {Netzer}, H.,
  {Sternberg}, A., {Mihos}, J.~C., {Baker}, A.~J., {Mazzarella}, J.~M., {Lord},
  S., {Sanders}, D.~B., {Stockton}, A., {Joseph}, R.~D., \& {Barnes}, J.~E.
  2009, \apjs, 182, 628

\bibitem[{{Wagg} {et~al.}(2012){Wagg}, {Pope}, {Alberts}, {Armus}, {Brodwin},
  {Bussmann}, {Desai}, {Dey}, {Jannuzi}, {Le Floc'h}, {Melbourne}, \&
  {Stern}}]{wagg12}
{Wagg}, J., {Pope}, A., {Alberts}, S., {Armus}, L., {Brodwin}, M., {Bussmann},
  R.~S., {Desai}, V., {Dey}, A., {Jannuzi}, B., {Le Floc'h}, E., {Melbourne},
  J., \& {Stern}, D. 2012, \apj, 752, 91

\bibitem[Zeimann et al.(2012)]{zeimann12} Zeimann, G.~R., 
Stanford, S.~A., Brodwin, M., et al.\ 2012, \apj, 756, 115 


\end{thebibliography}
